\documentclass[9pt,twocolumn,twoside]{osajnl}

\journal{jocn} 

\usepackage{caption} 
\usepackage{soul}
\usepackage{lipsum}
\usepackage{indentfirst}
\usepackage{booktabs}
\usepackage{enumitem}
\usepackage{tabularx}
\usepackage{multirow}
\usepackage{graphicx}
\usepackage{colortbl}
\usepackage{makecell} 
\usepackage{environ}         
\usepackage{etoolbox}        
\usepackage{pifont}
\usepackage{makecell}
\usepackage{array}
\usepackage{hhline}

\usepackage{graphicx}
\usepackage{wrapfig}
\usepackage[labelformat=simple]{subcaption}

\usepackage{float}
\usepackage{gensymb}
\usepackage{textpos}

\usepackage{amsmath}
\usepackage{amssymb}
\usepackage{amsfonts}
\usepackage{amsfonts, eqnarray, amsmath, amssymb}
\usepackage{amsthm}
\usepackage{newtxmath}
\usepackage{mathtools}
\usepackage{xfrac}
\usepackage{breqn}
\usepackage{bm}

\usepackage{algorithm}
\usepackage{algpseudocode}
\usepackage{etoolbox}
\algnewcommand{\LeftComment}[1]{\Statex \(\triangleright\) #1}

\makeatletter
\newcommand*{\algrule}[1][\algorithmicindent]{%
  \hspace*{.2em}
  \vrule 
  \hspace*{\dimexpr#1-.2em-.4pt}%
}

\newcommand{\StatePar}[1]{%
  \State\parbox[t]{\dimexpr\linewidth-\ALG@thistlm}{\strut #1\strut}%
}
\renewcommand{\ALG@beginalgorithmic}{\offinterlineskip}

\newcount\ALG@printindent@tempcnta
\def\ALG@printindent{%
  \ifnum \theALG@nested > 0
    \ifx\ALG@text\ALG@x@notext
    \else
      \unskip
      \ALG@printindent@tempcnta=1
      \loop
        \algrule[\csname ALG@ind@\the\ALG@printindent@tempcnta\endcsname]%
        \advance \ALG@printindent@tempcnta 1
        \ifnum \ALG@printindent@tempcnta<\numexpr\theALG@nested+1\relax
      \repeat
        \fi
    \fi
}
\patchcmd{\ALG@doentity}{\noindent\hskip\ALG@tlm}{\ALG@printindent}{}{\errmessage{failed to patch}}
\makeatother

\algrenewcommand\algorithmicend{\strut\textbf{end}}
\algrenewcommand\algorithmicdo{\strut\textbf{do}}
\algrenewcommand\algorithmicwhile{\strut\textbf{while}}
\algrenewcommand\algorithmicfor{\strut\textbf{for}}
\algrenewcommand\algorithmicforall{\strut\textbf{for all}}
\algrenewcommand\algorithmicloop{\strut\textbf{loop}}
\algrenewcommand\algorithmicrepeat{\strut\textbf{repeat}}
\algrenewcommand\algorithmicuntil{\strut\textbf{until}}
\algrenewcommand\algorithmicprocedure{\strut\textbf{procedure}}
\algrenewcommand\algorithmicfunction{\strut\textbf{function}}
\algrenewcommand\algorithmicif{\strut\textbf{if}}
\algrenewcommand\algorithmicthen{\strut\textbf{then}}
\algrenewcommand\algorithmicelse{\strut\textbf{else}}

\algrenewcommand\algorithmicrequire{\strut\textbf{Input:}}
\algrenewcommand\algorithmicensure{\strut\textbf{Output:}}

\let\oldState\State
\renewcommand{\State}{\oldState\strut}

\usepackage{array}
\usepackage{float}
\usepackage{textpos}
\usepackage{fancyhdr}
\usepackage[pscoord]{eso-pic}
\newcommand{\placetextbox}[3]{
  \setbox0=\hbox{#3}
  \AddToShipoutPictureFG*{
    \put(\LenToUnit{#1\paperwidth},\LenToUnit{#2\paperheight}){\vtop{{\null}\makebox[0pt][c]{#3}}}%
  }%
}%

\setboolean{shortarticle}{false}

\title{The Evolution of In-Premise Networks In the Era of Immersive Services: From Copper to Fiber-to-The Room [Invited Tutorial]}

\author{Elaine Wong}
\author{Sourav Mondal}

\affil{Dept of Electrical and Electronic Engineering University of Melbourne, VIC 3010, Australia}

\affil[*]{Corresponding author: ewon@unimelb.edu.au}

\placetextbox{0.5}{0.04}{This article is under review in IEEE/Optica Journal of Optical Communications and Networking. Copyright @ IEEE 2026.}%

\begin{abstract}
The telecommunications industry is transitioning towards an experience-driven era, shaped by immersive and interactive applications that place increasingly stringent demands on network performance. Over the past two and a half decades, passive optical network (PON) based Fiber-To-The-Home (FTTH) networks have been central to delivering high-capacity broadband access to homes and businesses. However, the growing performance gap between the access and in-premise indoor networks creates an impetus for driving fiber into home and buildings.  Complementing rather than replacing copper cabling solutions and wireless access technologies, e.g. WiFi, the Fiber-to-the-Room (FTTR) solution is set to accelerate over the next few years. This article examines the immersive applications, in particular collaborative extended reality (XR), driving this shift.  It discusses the multi-modal traffic characteristics of collaborative XR, Quality-of-Experience (QoE) requirements, corresponding implications on network performance along with emerging solutions including robust resource allocation aimed at meeting QoE requirements.  In particular, the article presents a novel resource allocation framework that combines coarse-grained traffic forecasting and intelligent in-premise roaming. The framework enables flexible selection of artificial intelligence models depending on the computational capability of each network segment, namely FTTH, FTTR, and WiFi. Numerical evaluations of a representative FTTH-FTTR-WiFi network architecture show that long-horizon, multi-step-ahead traffic forecasts can reduce inference frequency and enable coordinated resource allocation and network-orchestrated seamless user mobility across rooms that satisfies the stringent QoE requirements of XR collaborations. This article concludes by highlighting opportunities to further amortize investment in FTTR infrastructure by extending its role beyond connectivity. Through leveraging the same deployed optical infrastructure to support multiple services and capabilities, FTTR offers opportunities to improve infrastructure utilization, enable new service and revenue models, and maximize the value of fiber deployment within the premises.
\end{abstract}

\setboolean{displaycopyright}{false} 

\begin{document}

\maketitle

\section{Introduction}
Today, telecommunication networks are transitioning from a communication-oriented era based on voice, video and data, towards a cognitive- and eventually experience-oriented era underpinned by high-capacity connectivity, cloud/edge computing, artificial intelligence (AI), computer vision, advanced sensing, and immersive technologies \cite{WongJOCNTutorial2023, WongOFC2026}. We are now witnessing the acceleration of physical and digital convergence through collaborative extended reality (XR), tactile and haptic interactions, and ultimately holographic communications \cite{ind_metav, h2m_sourav}. In turn, this is transforming how humans interact with information, services, and each other. \par
3GPP defines Extended Reality (XR) as all "real and virtual combined environments and associated human-machine interactions generated by computer technology and wearables as extended reality" \cite{3GPP_TR26928}. It is an umbrella term encompassing three immersive technologies, namely augmented reality (AR), mixed reality (MR), and virtual reality (VR), which span the continuum from the physical world to fully virtual environments.  While the gaming and entertainment industries are driving XR adoption today, immersive applications based on XR are gaining momentum across numerous industry sectors \cite{xr_train, xr_survey, Rao2016Potholes}.  These include remote operation of machinery, collaborative maintenance, worker training, digital twin visualization, and human-robot collaboration in industrial automation and Industry 5.0; remote diagnosis, tele-rehabilitation, surgical training, and remote robotic intervention in healthcare and rehabilitation; immersive and collaborative learning and training environments; virtual showrooms and product demonstrations in commerce and retail; environmental awareness, navigation and mobility enhancement in assistive technology; and virtual tourism experiences and digitally enhanced exhibitions within the tourism sector. The immersive capability and diversity of these applications demonstrate a change in the requirements imposed on the underlying network, whereby connectivity and peak data rate alone are no longer sufficient because application performance becomes increasingly dependent on the ability of the end-to-end network to simultaneously provide high throughput, low latency, high reliability, and consistent service delivery. Along with advances in optical core and metro networks, today's Fiber-to-the-Home (FTTH) and in-premise networks must remain fit for purpose to support immersive services. \par
FTTH networks based on passive optical network (PON) technologies have undergone rapid capacity growth. Since the early 2000s, PON technology has evolved through multiple generations, scaling from 2.5-Gb/s G-PON \cite{S1_ITU2008} to 10-Gb/s XG(S)-PON \cite{S2_ITU2012, S2_ITU2016}, 40-Gb/s NG-PON2 \cite{S2_ITU2015}, and, more recently, 50G-PON \cite{S2_ITU2021}, with emerging 200-Gb/s-class PON technologies \cite{S2_ITUVHSP} now under consideration.  This evolution provides an increasingly high-capacity optical path between the central office (CO) and the residential or enterprise gateway, but one that is not necessarily matched within the premises in which connectivity is predominantly wireless technology based and may therefore not consistently satisfy user experience requirements.

\begin{figure}[!t]
\centering
\includegraphics[width=\columnwidth]{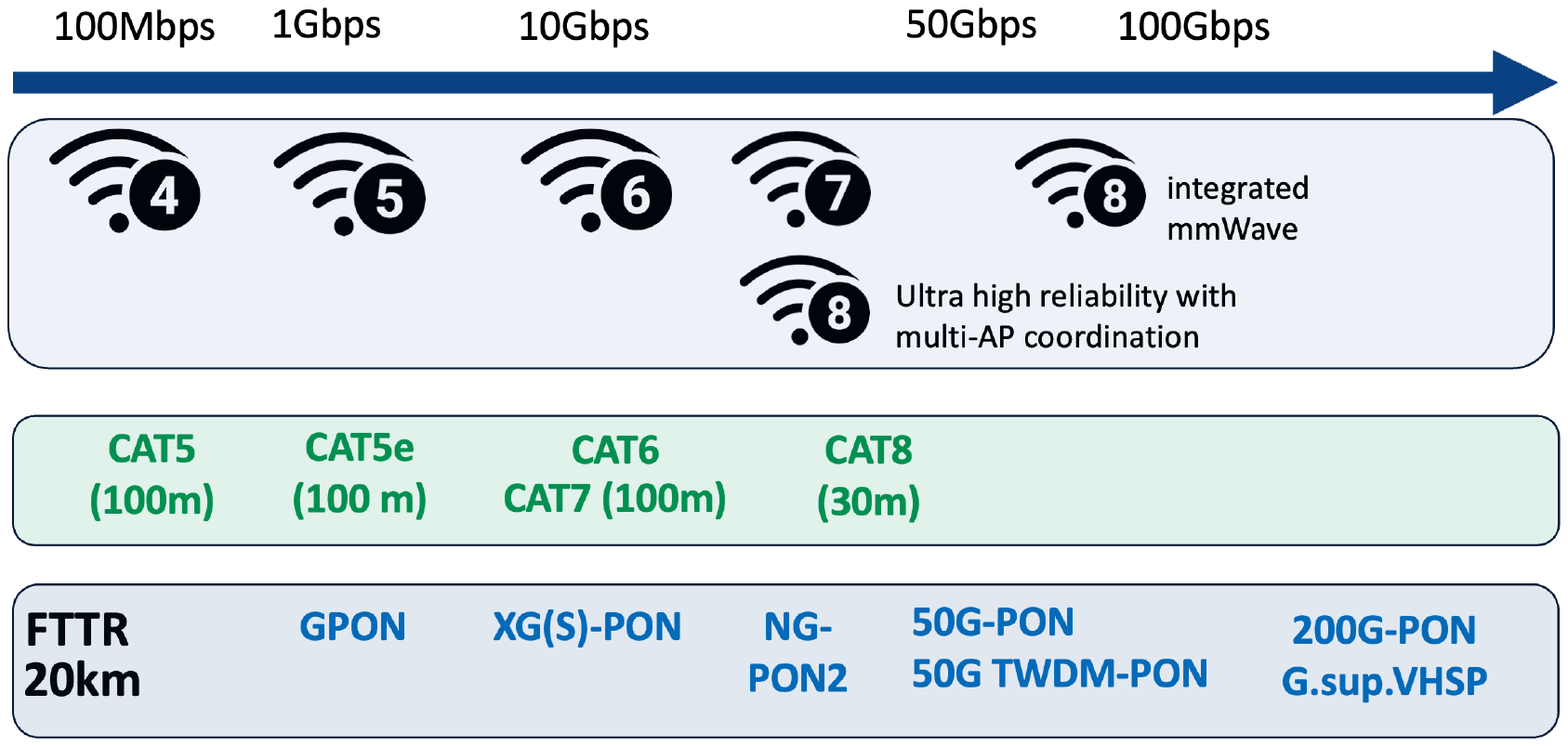}
\caption{Evolution of in-premise access technologies.  Adapted from \cite{WongOFC2026}.}
\label{AccessTechnologies}
\end{figure}
\setlength{\textfloatsep}{5pt}

According to \cite{wbba2023}, the number of connected devices within a household is projected to increase from approximately 13 in 2020 to 44 by 2030, while the device ecosystem is evolving from predominantly smartphones, personal computers, and 2D displays towards XR headsets, home robotics, AI companions, ambient sensors, and live 3D displays. Thus, in-premise networks must transition from supporting connected devices to supporting connected experiences, in which multiple high-bandwidth and latency-sensitive immersive applications may operate concurrently. The unprecedented growth in both the number and capability of connected in-home devices is expected to result in a corresponding increase in traffic demand, with representative data-rate requirements progressing from approximately 50 Mb/s for 4K video and 300 Mb/s for 8K video to 1.5 Gb/s for immersive XR collaboration and 5.5 Gb/s for live 3D video. Importantly, these data rates must be sustained concurrently with stringent latency, reliability, and connectivity requirements. Hence, the growing performance gap between the external high-capacity FTTH access network and the internal in-premise network over which immersive services are ultimately delivered, warrants attention.

\begin{figure*}[!t]
\centering
\includegraphics[width=0.9\textwidth]{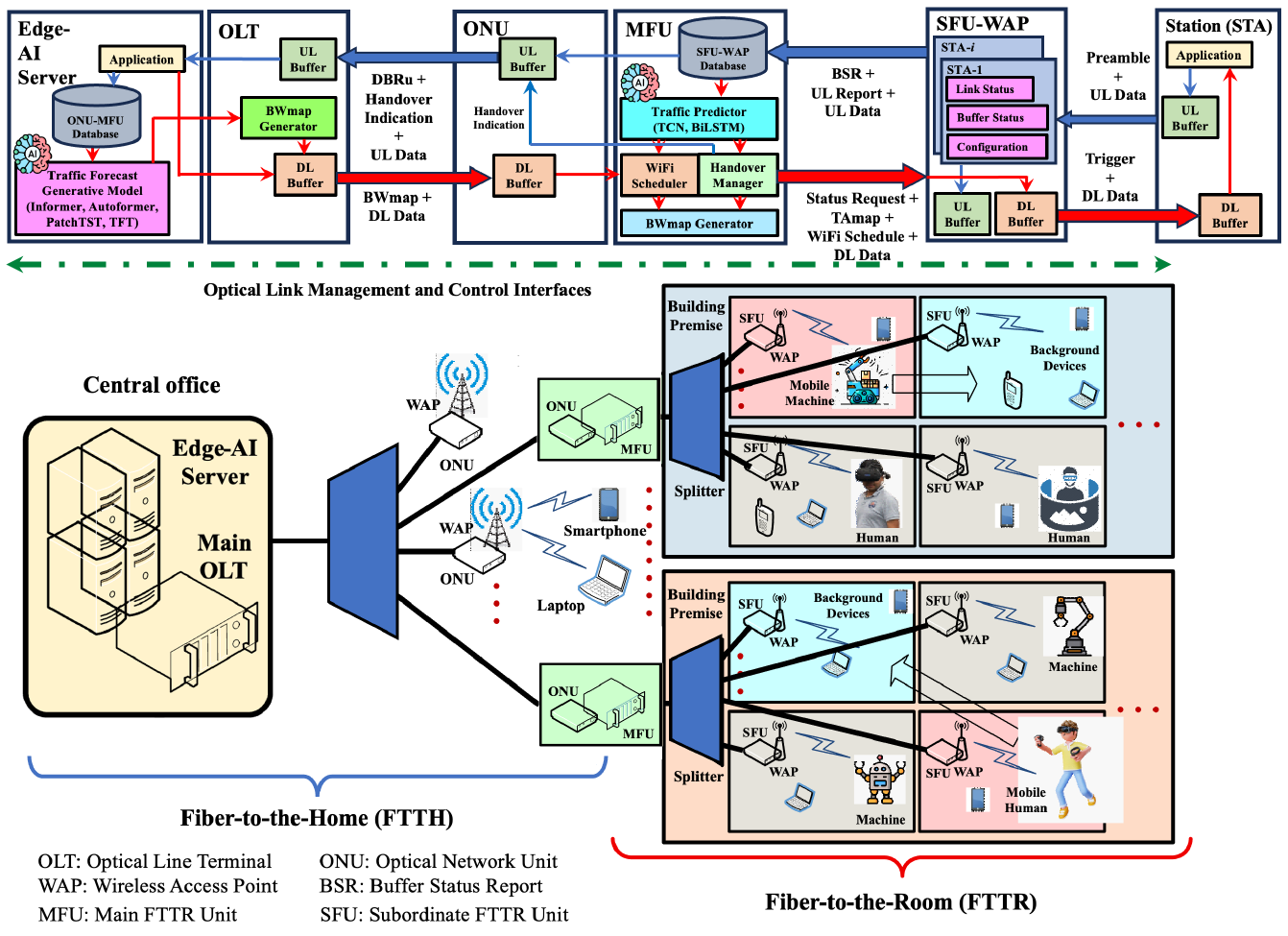}
\caption{The multi-tier integrated FTTH-FTTR-WiFi network architecture embedded with Gen-Pred AI framework. This architecture enables proactive resource allocation and intelligent roaming to satisfy the QoE requirements of multi-modal traffic during XR collaborations.}
\label{arch1}
\end{figure*}
\setlength{\textfloatsep}{1pt}

Fiber-to-the-Room (FTTR) addresses this discontinuity by extending fiber from the FTTH gateway deep into the premises, typically over distances of up to approximately 100 m, and using fiber as the high-capacity backhaul for distributed short-range wireless access \cite{itu_fttr, ITUTG9940, ITUTG9942, Polan_2017}. FTTR is not intended to replace WiFi but rather to complement its strengths: fiber provides capacity, reach, and stable backhaul connectivity between rooms, while advanced WiFi provides mobility and wireless connectivity over the final several metres to end devices. This architecture is particularly relevant to immersive applications, for which high peak wireless PHY rates alone are insufficient if the corresponding backhaul cannot maintain comparable capacity and predictable performance.

In this article, we aim to examine the evolution of in-premise networks towards fiber-based architectures capable of supporting emerging collaborative XR applications. In particular, we discuss the perceptual and network requirements imposed by these applications and their multi-modal traffic. Recognizing that satisfying these requirements depends not only on increasing network capacity but also on efficiently coordinating resources across the FTTH, FTTR, and WiFi segments, we present a generative-predictive (Gen-Pred) AI framework for proactive resource allocation and intelligent in-premise roaming. The framework exploits long-horizon, multi-step-ahead traffic forecasting to anticipate resource demand, while flexibly selecting AI models according to the computational capabilities and forecasting requirements of different network segments. Specifically, predictive traffic forecasting is performed within the FTTR segment for fine-grained resource allocation, while Transformer-based generative forecasting is employed at the edge for coarse-grained resource allocation within the external FTTH network. These mechanisms are further integrated with network-orchestrated intelligent roaming to maintain the stringent QoE requirements of collaborative XR applications as users move throughout the premises. Finally, acknowledging the higher initial investment associated with deploying fiber deep into homes and buildings, we discuss opportunities to further amortize the FTTR infrastructure through an emerging FTTR + "x" model, whereby the same optical infrastructure can support additional "x" services and capabilities beyond connectivity.

\section{Extending Fiber into the Premises}
\subsection{Evolution of Indoor Technologies}
Fig. \ref{AccessTechnologies} charts the broader evolution of in-premise wireless and copper-based access technologies and compares the role of FTTR within this trajectory. In-premise networks have traditionally used copper-based Ethernet cabling, which has evolved substantially in transmission capacity to 40 Gbps based on the Category 8 (CAT8) standard \cite{IEEE8023bq}. However, capacity improvements are accompanied by increasingly stringent distance limitations, with CAT8 supporting its highest data rates over only approximately 30 m. In addition, copper-based Ethernet offers limited upgrade flexibility compared with fiber. Supporting successive generations of multi-gigabit services may therefore require replacement of the installed copper cabling infrastructure, increasing both the cost and complexity of network upgrades. In comparison, optical fiber possesses several benefits that make it particularly attractive as an indoor cabling solution. It offers immunity to electromagnetic interference, low attenuation, high reliability, and long operational lifetimes when compared to copper-based solutions \cite{Polan_2017}.

 WiFi standards have also continued to evolve. Theoretically, WiFi-6/6E and WiFi 7 can support up to maximum datarates of $\sim$9.6 Gbps \cite{ieee80211ax} and $\sim$46 Gbps ~\cite{IEEE80211be2024}, respectively.  There is also ongoing development of WiFi-8, including IEEE 802.11bn (operating below 7 GHz) \cite{IEEE80211bn} and IEEE 802.11bq (41–70 GHz) \cite{IEEE80211bq} with targeted maximum data rates of approximately 30 Gb/s and 59 Gb/s, respectively. As importantly, each WiFi standard introduces significant functional and architectural enhancements. For example, WiFi-6 adopts uplink scheduling and orthogonal frequency-division multiple access (OFDMA) to enhance spectrum efficiency and reduce latency \cite{ieee80211ax}. Building on this, WiFi-7 introduces multi-link operation (MLO), enabling devices to aggregate multiple frequency bands for a higher throughput ~\cite{IEEE80211be2024}. IEEE 802.11bn introduces coordinated multi-WAP transmission to minimize interference and ensure deterministic, ultra-low-latency communication~\cite{IEEE80211bn}, while IEEE 802.11bq introduces integrated millimeter-wave (mmWave) with conventional sub-7.25-GHz operation in order to achieve high capacity mmWave links over short distances \cite{IEEE80211bq}.

Exploiting mature XG(S)-PON technology and, in the future, the 50G-PON technology, FTTR can provide a future-proof optical backhaul for interconnecting densely distributed Wireless Access Points (WAPs) and preserving wireless performance gains throughout the premises.  Beyond high capacity and low attenuation, resilience and service continuity can also be incorporated using protection and switching mechanisms previously explored for FTTH \cite{Nadarajah2006Protection, Wong2014Survivable, Yu2025ConceptDrift}.  FTTR is therefore complementary to, rather than a replacement for WiFi technologies. This complementarity becomes particularly important with the introduction of mmWave operation, where higher propagation and penetration losses necessitate shorter wireless links and, consequently, denser WAP deployments. FTTR and WiFi can therefore evolve in tandem, with fiber providing a future-proof in-premise backhaul infrastructure capable of supporting successive generations of wireless access without requiring replacement of the underlying optical cabling.

\subsection{FTTR Architecture}
Building on the preceding discussion, Fig. \ref{arch1} illustrates a representative integrated FTTH-FTTR-WiFi network architecture. An optical line terminal (OLT) at the central office of an external FTTH network is connected to the main fiber unit (MFU) of an FTTR in-premise network \cite{ITUTG9940}. In this representative architecture, the FTTH and FTTR networks are deployed using 50G-PON and XG(S)-PON technologies, respectively. The MFU is the residential gateway of each home and serves two purposes: as an optical network unit (ONU) to the external FTTH network, and as an OLT for the FTTR in-premise network. Within the FTTR network, the MFU is connected to a series of subordinate fiber units (SFUs) through a point-to-multi-point distribution network.  Further, by leveraging new control and management interfaces, the MFU is able to coordinate resource allocation, intelligent user roaming across rooms, interference mitigation and power optimization~\cite{fttr_zte}. 

In turn, each SFU is integrated with a WAP in each room to provide short-range (approx. 10 meters) wireless connectivity to user devices or stations (STAs) within that room ~\cite{fttr_mag}. Interference with neighboring SFU-WAPs is alleviated as each room is its own unique collision domain. Moreover, with each SFU-WAP serving only a small area, network reliability and stability are improved as dependence on long-range wireless links that suffer from structural losses is reduced. Overall, when combined with short-reach, high-capacity advanced wireless technologies, the passive fiber infrastructure of FTTR serves as a future-proof and reliable backhaul to complement the mobility and device connectivity of the last 10 meters. \par

\begin{figure*}[!t]
\centering
\includegraphics[width=0.9\textwidth,keepaspectratio]{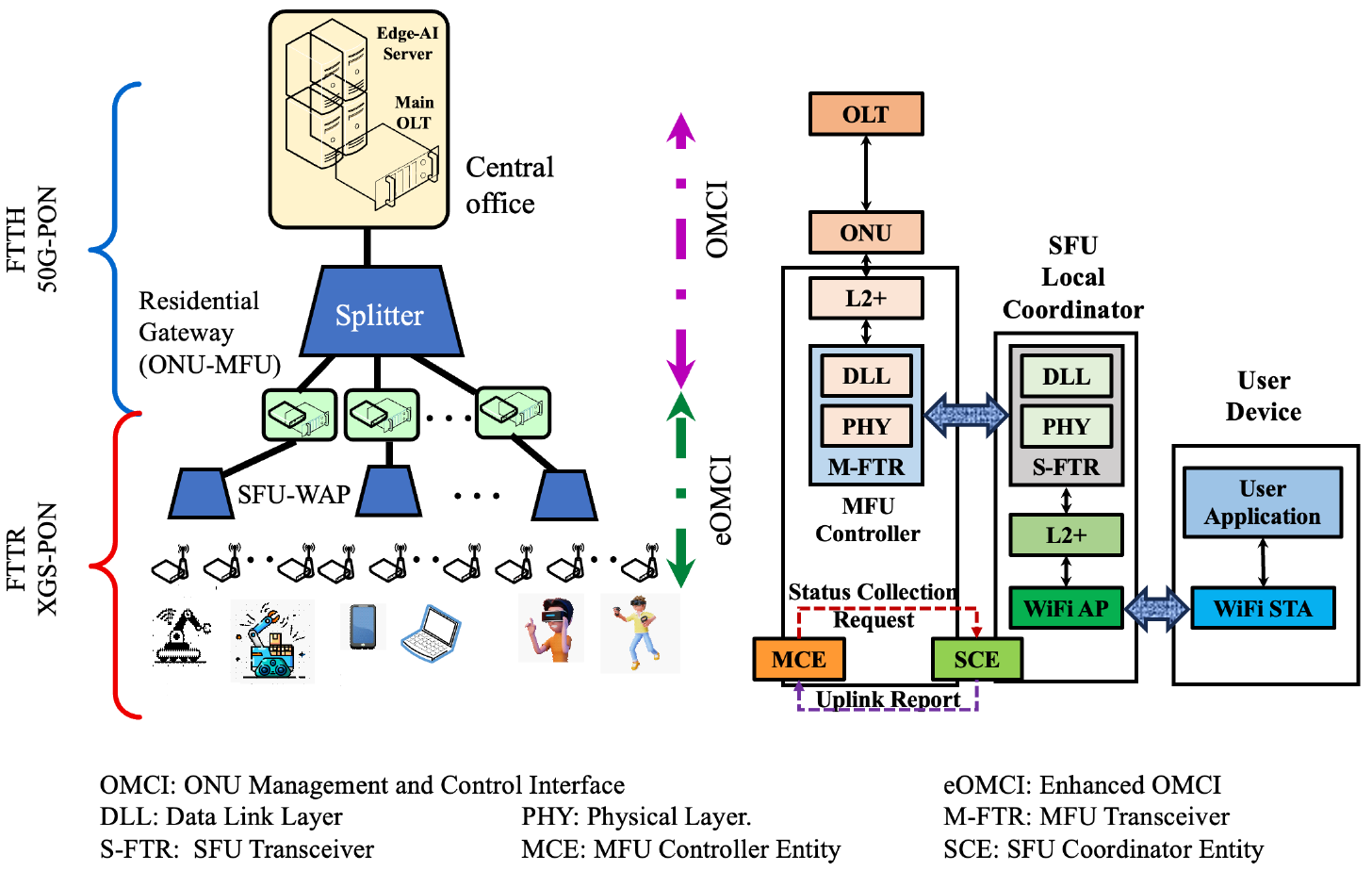}
\caption{Management and control interfaces of the multi-tier integrated FTTH-FTTR-WiFi architecture.}
\label{figemoci}
\end{figure*}
\setlength{\textfloatsep}{0pt}
\subsection{Deployment Considerations}
While the concept of extending fiber into premises has existed since a decade ago, e.g. \cite{NokiaPOL2015}, \cite{TellabsPOL2015}, and \cite{Polan_2017}, FTTR has only transitioned to large-scale commercial deployment since the early 2020s. According to \cite{LightCountingFTTR2024}, the number of FTTR subscribers worldwide is forecast to reach nearly 200 million by 2029, with China accounting for approximately 94\% of the global market. As the initial large-scale deployment in China approaches maturity, subscribers in the rest of the world are forecast to grow at a substantially higher compound annual growth rate (CAGR) of 112\% (2023-2029), compared with 57\% in China. This reflects an emerging second phase of FTTR deployment as operators across Asia-Pacific, Europe, the Middle East and the Americas begin commercial rollouts.

Despite its advantages, FTTR faces several practical deployment challenges. Compared with low-cost copper and WiFi alternatives that can often be self-installed, the higher upfront cost of FTTR, arising from both capital expenditure and the need for specialized installation, continues to limit widespread adoption \cite{LightCountingFTTR2024}. Retrofitting existing or older buildings may introduce additional complexity due to regulatory and building approval requirements.  The visibility of newly installed fiber can also present aesthetic concerns where concealed cabling or conduits are not available. Nonetheless, recent advances in pre-connectorized fiber, invisible fiber cables with ultra-low bending radius, and adhesive bonding installation methods are addressing these challenges by simplifying indoor fiber installation deployment while reducing its visual impact \cite{HolightFTTR2024}, \cite{HuaweiFTTR}. More importantly, FTTR provides a future-ready in-premises infrastructure upon which new services and capabilities can be introduced, providing further motivation for its deployment, as discussed in Section 6.
\subsection{Centralized Management and Control} 
To facilitate the control and management of the FTTR network, the standard ONU Management and Control Interface (OMCI) is extended to the enhanced OMCI (eOMCI) as described in ITU-T G.Sup.eOMCI \cite{ETSI_GR_F5G032}. As illustrated in Fig. \ref{figemoci}, eOMCI allows the transparent forwarding of control messages across the OLT–MFU and MFU–SFU interfaces.  Each MFU therefore acts as a proxy OLT for the SFUs, employing an internal OMCI-based management interface for configuration, firmware upgrade, and fault supervision, along with a complementary control interface to support real-time operations \cite{ITUTG9940}. Additionally, WiFi coordination messages are exchanged via the WLAN management and control interface (WMCI). In this control plane, the MFU controller entity (MCE) sends \emph{status collection requests} to all the SFU coordinator entities (SCEs). In response, the SCEs send \emph{uplink reports} to the MCE, containing airtime utilization, channel conditions, received signal strengths, interference levels, and buffer status reports (BSRs) of all the associated STAs. Using this information, the MCE formulates coordinated strategies for optimal resource allocation and seamless in-premise handovers. The corresponding control commands are then transmitted to the SCEs who execute them locally over both the FTTR and WiFi domains \cite{ITUTG9942}. Note that both the eOMCI and WMCI control messages are multiplexed with user data for transmission as a Type-4 T-CONT (TC-4) flow utilizing the same wavelength(s). To ensure the reliability of this in-band signaling, different IEEE 802.1p Class-of-Service (CoS) priorities can be assigned internally at MFU/SFUs, e.g., the highest priority (CoS-7) for eOMCI and WMCI control messages, followed by HMD orientation, control, and haptics (CoS-6), video (CoS-5), and background traffic (CoS-1). Such differentiation ensures that network coordination and latency-critical immersive traffic are protected even under periods of high in-premises network load.

Crucially, the centralized FTTR-WiFi integrated architecture can be leveraged to enable intelligent network-orchestrated in-premise roaming \cite{fttr_coop}. In  \cite{fttr_jocn}, seamless inter-WAP handovers between 4$\times$4~m$^2$ rooms were experimentally demonstrated, achieving $\leq$20 msec latency. In  \cite{fttr_ofc_25}, inter-WAP handovers spaced 6~meters apart were demonstrated in a field-trial, achieving  $\leq$35 msec latency.  In \cite{fttrb_ofc}, we showed that efficient network-orchestrated handovers between WAPs in 10$\times$10~m$^2$ rooms can reduce air-interface latency by up to 90\%.\par

\section{Perceptual and Network Requirements of Immersive XR} \label{sec3}
Historically, access and in-premise networks have supported early broadband services such as of VoIP, multimedia streaming, web browsing and email. Information flowed predominantly in one direction from the central office to the end users in homes and buildings. Although video streaming significantly increased bandwidth requirements, user interaction remained relatively limited. With the rise of immersive applications such as collaborative XR, these networks must also satisfy perceptual QoE requirements associated with human cognition, in addition to capacity requirements. \par
\subsection{Quality-of-Experience (QoE)}
\begin{figure}[!t]
\centering
\includegraphics[width=\columnwidth,keepaspectratio]{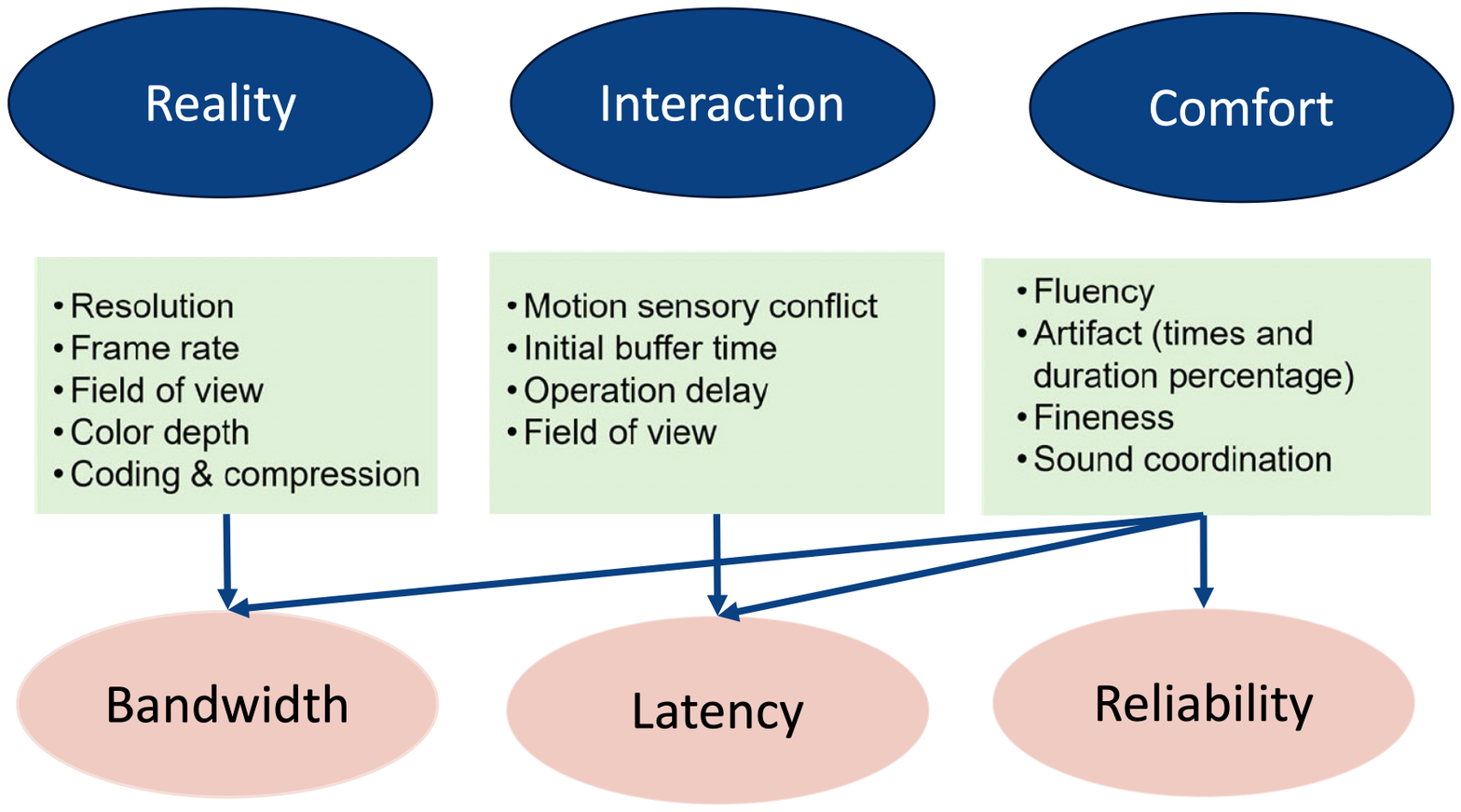}
\caption{Quality-of-Experience (QoE) dimensions and implications on network performance metrics.  Adapted from \cite{Xu2026VLCXR}.}
\label{QoE}
\end{figure}

Quality of Experience (QoE) refers to the degree to which a communication system can create, sustain, and synchronize immersive environments that resemble and respond to the physical world. QoE is governed by three principal experiential dimensions, namely the sense of reality, the sense of interaction, and the sense of comfort \cite{Xu2026VLCXR}, as illustrated in Fig. ~\ref{QoE}.  The sense of reality largely stems from how convincingly the system can reproduce visual and auditory stimuli that align with human perception. This depends on factors such as display resolution, frame rate, color accuracy, and the efficiency of video compression. Maintaining visual fidelity therefore requires communication links with sufficient bandwidth to sustain high definition, real-time content delivery. Similarly, the sense of interaction depends on how quickly and precisely the system responds to user inputs. In XR systems, rendering and computation are partially or entirely offloaded to cloud or edge servers. This distributed processing makes network latency a key determinant of responsiveness. Even small transmission delays or jitter can cause mismatches between user motion and visual updates, resulting in discomfort or motion sickness. Finally, the sense of comfort relates to the smoothness and stability of the entire XR session. Frame drops, image tearing, and synchronization errors disrupt the continuity of experience and reduce overall satisfaction. To ensure fluid and enjoyable immersion, the network must guarantee sufficient throughput, low latency, and stability under dynamic conditions.  

\setlength{\textfloatsep}{5pt}
\subsection{Network requirements} 
The three experiential dimensions demonstrate that user experience in XR is not governed by just hardware capability and software algorithms, but also tightly coupled to the performance of the communication network that supports it.  For in-premise networks, this means moving beyond providing legacy best-effort networks toward high-capacity, low-latency solutions.

Unsurprisingly, it follows that deploying XR applications over existing communication networks, especially across large geographic spans and in-premise environments, presents several complex network challenges. A typical collaborative XR session involves the continuous and concurrent exchange of multiple traffic modalities.  Table~\ref{table1} summarizes the \emph{fair}, \emph{comfortable}, and \emph{ideal QoE} requirements associated with these modalities, including ultra high-resolution (4K, 8K, and 16K) video, head-movement data represented by head-mounted display (HMD) or optical HMD (OHMD) orientation data, tactile control signals, and haptic feedback data. XR video imposes the highest bandwidth requirement. Increasing spatial resolution and frame rate towards those required to approximate human visual perception results in extremely high source data rates, reaching several Tb/s for uncompressed video. Recent frame compression techniques exploiting temporal redundancy between consecutive frames such as high efficiency video coding (H.265/HEVC) and versatile video coding (H.266/VVC) \cite{vvc} can significantly reduce the datarate when the frames are transmitted with high-reliability. For smoothness and responsiveness of the visuals, modern HMDs/OHMDs typically support a frame refresh rate of  90-120~Hz \cite{deb_vr}. Even with compression, achieving ideal XR QoE requires at least 1.5 Gb/s per video stream. In collaborative scenarios involving multiple users, viewpoints, or remotely generated video streams, the aggregate bandwidth requirement can therefore rapidly increase to multiple Gb/s. 

It is also evident that the interactive nature of immersive XR imposes stringent timing constraints across multi-modal traffic streams, with an end-to-end latency bound of approximately 10 ms required for ideal QoE. Compounding this challenge, maintaining perceptual continuity and seamless human–machine interaction requires synchronisation across multiple traffic modalities, such that these stringent latency requirements must be satisfied concurrently despite their differing traffic characteristics.
To satisfy these requirements, application layer and network layer collaborative approaches like network-aware adaptive XR video streaming \cite{xr_5g1} and XR QoE-aware scheduling \cite{xr_5g2} have been proposed. Moreover, in  \cite{vr_mpg} a framework for immersive multi-user game was proposed whereby XR video frames of each player according to their field-of-view (FOV) and game status are pre-rendered for low-latency transmission. In \cite{lstm_xr}, the proposed long short-term memory (LSTM)-based FOV prediction method is shown to reduce bandwidth demand by 61\% by transmitting FOV tiles at higher quality. However, such solutions alone cannot satisfy XR collaboration QoE requirements without sufficient network capacity and deterministic connectivity.
\par
\begin{table*}[!t]
\caption{Summary of Immersive XR Modalities: QoE and QoS requirements and traffic characteristics \cite{xr_std,xr_iotj}}
\label{table1}
\centering
%
\setlength\arrayrulewidth{1.5pt}
\begin{tabular}{
>{\columncolor[HTML]{EFEFEF}}c 
>{\columncolor[HTML]{EFEFEF}}c cccccc}
\hline
\multicolumn{8}{c}{\cellcolor[HTML]{C0C0C0}\textbf{Traffic Profile}} \\ \hline
\multicolumn{2}{c|}{\cellcolor[HTML]{EFEFEF}} &
  \multicolumn{1}{c|}{\cellcolor[HTML]{EFEFEF}} &
  \multicolumn{1}{c|}{\cellcolor[HTML]{EFEFEF}} &
  \multicolumn{1}{c|}{\cellcolor[HTML]{EFEFEF}} &
  \multicolumn{1}{c|}{\cellcolor[HTML]{EFEFEF}} &
  \multicolumn{1}{c|}{\cellcolor[HTML]{EFEFEF}} &
  \cellcolor[HTML]{EFEFEF} \\
\multicolumn{2}{c|}{{\cellcolor[HTML]{EFEFEF}\textbf{Traffic Type}}} &
  \multicolumn{1}{c|}{{\cellcolor[HTML]{EFEFEF}\textbf{\begin{tabular}[c]{@{}c@{}}Frame\\ Rate\end{tabular}}}} &
  \multicolumn{1}{c|}{{\cellcolor[HTML]{EFEFEF}\textbf{\begin{tabular}[c]{@{}c@{}}Avg.\\ Datarate\end{tabular}}}} &
  \multicolumn{1}{c|}{{\cellcolor[HTML]{EFEFEF}\textbf{Bandwidth}}} &
  \multicolumn{1}{c|}{{\cellcolor[HTML]{EFEFEF}\textbf{\begin{tabular}[c]{@{}c@{}}Packet\\ Arrival Time\end{tabular}}}} &
  \multicolumn{1}{c|}{{\cellcolor[HTML]{EFEFEF}\textbf{\begin{tabular}[c]{@{}c@{}}Packet\\ Size (Byte)\end{tabular}}}} &
  {\cellcolor[HTML]{EFEFEF}\textbf{\begin{tabular}[c]{@{}c@{}}Latency\\ Bound\end{tabular}}} \\ \hline
\multicolumn{1}{c|}{\cellcolor[HTML]{EFEFEF}} &
  \multicolumn{1}{c|}{\cellcolor[HTML]{EFEFEF}} &
  \multicolumn{1}{c|}{} &
  \multicolumn{1}{c|}{} &
  \multicolumn{1}{c|}{} &
  \multicolumn{1}{c|}{} &
  \multicolumn{1}{c|}{} &
   \\
\multicolumn{1}{c|}{\cellcolor[HTML]{EFEFEF}} &
  \multicolumn{1}{c|}{{\cellcolor[HTML]{EFEFEF}\textbf{\begin{tabular}[c]{@{}c@{}}Fair QoE\\ {[}4K{]}\end{tabular}}}} &
  \multicolumn{1}{c|}{{*}{\begin{tabular}[c]{@{}c@{}}60\\ fps\end{tabular}}} &
  \multicolumn{1}{c|}{{\begin{tabular}[c]{@{}c@{}}40-90\\ Mbps\end{tabular}}} &
  \multicolumn{1}{c|}{{\begin{tabular}[c]{@{}c@{}}$\geq$ 260\\ Mbps\end{tabular}}} &
  \multicolumn{1}{c|}{} &
  \multicolumn{1}{c|}{} &
  {$\leq$20 msec} \\ \cline{2-5} \cline{8-8} 
\multicolumn{1}{c|}{\cellcolor[HTML]{EFEFEF}} &
  \multicolumn{1}{c|}{\cellcolor[HTML]{EFEFEF}\textbf{\begin{tabular}[c]{@{}c@{}}Comfortable\\ QoE {[}8K{]}\end{tabular}}} &
  \multicolumn{1}{c|}{\begin{tabular}[c]{@{}c@{}}{[}60,90{]}\\ fps\end{tabular}} &
  \multicolumn{1}{c|}{\begin{tabular}[c]{@{}c@{}}100-360\\ Mbps\end{tabular}} &
  \multicolumn{1}{c|}{\begin{tabular}[c]{@{}c@{}}$\geq$ 1\\ Gbps\end{tabular}} &
  \multicolumn{1}{c|}{} &
  \multicolumn{1}{c|}{} &
  $\leq$15 msec \\ \cline{2-5} \cline{8-8} 
\multicolumn{1}{c|}{{\cellcolor[HTML]{EFEFEF}\textbf{Video}}} &
  \multicolumn{1}{c|}{\cellcolor[HTML]{EFEFEF}\textbf{\begin{tabular}[c]{@{}c@{}}Ideal QoE\\ {[}16K{]}\end{tabular}}} &
  \multicolumn{1}{c|}{\begin{tabular}[c]{@{}c@{}}{[}60,90,120{]}\\ fps\end{tabular}} &
  \multicolumn{1}{c|}{\begin{tabular}[c]{@{}c@{}}400-800\\ Mbps\end{tabular}} &
  \multicolumn{1}{c|}{\begin{tabular}[c]{@{}c@{}}$\geq$ 1.5\\ Gbps\end{tabular}} &
  \multicolumn{1}{c|}{{\begin{tabular}[c]{@{}c@{}}Truncated Gaussian,\\ Mean = $\frac{1}{\text{fps}}$,\\ SD = 2 msec\end{tabular}}} &
  \multicolumn{1}{c|}{{\begin{tabular}[c]{@{}c@{}}Truncated Gaussian,\\ Mean = $\frac{\text{Avg. Datarate}}{(\text{fps} \times 8)}$,\\ SD = 10.5\% of Mean\end{tabular}}} &
  $\leq$10 msec \\ \hline
\multicolumn{2}{c|}{\cellcolor[HTML]{EFEFEF}\textbf{\begin{tabular}[c]{@{}c@{}}HMD \\ Orientations\end{tabular}}} &
  \multicolumn{1}{c|}{66 fps} &
  \multicolumn{1}{c|}{\begin{tabular}[c]{@{}c@{}}0.2-0.3\\ Mbps\end{tabular}} &
  \multicolumn{1}{c|}{\begin{tabular}[c]{@{}c@{}}$\geq$ 50\\ Mbps\end{tabular}} &
  \multicolumn{1}{c|}{\begin{tabular}[c]{@{}c@{}}Gamma,\\ Mean = 15 msec,\\ SD = 2 msec\end{tabular}} &
  \multicolumn{1}{c|}{100-500} &
  $\leq$10 msec \\ \hline
\multicolumn{2}{c|}{\cellcolor[HTML]{EFEFEF}\textbf{\begin{tabular}[c]{@{}c@{}}Tactile \\ Control\end{tabular}}} &
  \multicolumn{1}{c|}{100 fps} &
  \multicolumn{1}{c|}{\begin{tabular}[c]{@{}c@{}}1-2 \\ Mbps\end{tabular}} &
  \multicolumn{1}{c|}{\begin{tabular}[c]{@{}c@{}}$\geq$ 200\\ Mbps\end{tabular}} &
  \multicolumn{1}{c|}{\begin{tabular}[c]{@{}c@{}}Generalized Pareto,\\ Mean = 10 msec,\\ SD = 2 msec\end{tabular}} &
  \multicolumn{1}{c|}{1000-1500} &
  1-10 msec \\ \hline
\multicolumn{2}{c|}{\cellcolor[HTML]{EFEFEF}\textbf{\begin{tabular}[c]{@{}c@{}}Haptic \\ Feedback\end{tabular}}} &
  \multicolumn{1}{c|}{100 fps} &
  \multicolumn{1}{c|}{\begin{tabular}[c]{@{}c@{}}1-2\\ Mbps\end{tabular}} &
  \multicolumn{1}{c|}{\begin{tabular}[c]{@{}c@{}}$\geq$ 200\\ Mbps\end{tabular}} &
  \multicolumn{1}{c|}{\begin{tabular}[c]{@{}c@{}}Generalized Pareto,\\ Mean = 10 msec,\\ SD = 2 msec\end{tabular}} &
  \multicolumn{1}{c|}{1000-1500} &
  1-10 msec \\ \hline
\end{tabular}
\end{table*}
\setlength{\textfloatsep}{1pt}


\begin{table*}[!t]
\centering 
\caption{\bf Summary of multi-modal traffic in immersive XR applications}
\label{table2}
\centering
%
\setlength\arrayrulewidth{1.5pt}
\begin{tabular}{
>{\columncolor[HTML]{EFEFEF}}c cccccc}
\hline
\multicolumn{7}{c}{\cellcolor[HTML]{CBCEFB}\textbf{Immersive XR Collaboration}} \\ \hline
\multicolumn{1}{c|}{\cellcolor[HTML]{EFEFEF}} &
  \multicolumn{1}{c|}{\cellcolor[HTML]{ECF4FF}} &
  \multicolumn{1}{c|}{\cellcolor[HTML]{ECF4FF}} &
  \multicolumn{1}{c|}{\cellcolor[HTML]{ECF4FF}} &
  \multicolumn{1}{c|}{\cellcolor[HTML]{ECF4FF}} &
  \multicolumn{2}{c}{\cellcolor[HTML]{ECF4FF}\textbf{H2M}} \\ \cline{6-7} 
\multicolumn{1}{c|}{{\cellcolor[HTML]{EFEFEF}\textbf{Traffic Type}}} &
  \multicolumn{1}{c|}{{\cellcolor[HTML]{ECF4FF}\textbf{VR}}} &
  \multicolumn{1}{c|}{{\cellcolor[HTML]{ECF4FF}\textbf{AR}}} &
  \multicolumn{1}{c|}{{\cellcolor[HTML]{ECF4FF}\textbf{MR}}} &
  \multicolumn{1}{c|}{{\cellcolor[HTML]{ECF4FF}\textbf{H2H}}} &
  \multicolumn{1}{c|}{\cellcolor[HTML]{ECF4FF}\textbf{Human}} &
  \cellcolor[HTML]{ECF4FF}\textbf{Machine} \\ \hline
\multicolumn{1}{c|}{{\cellcolor[HTML]{EFEFEF}\textbf{Video}}} &
  \multicolumn{1}{c|}{{DL}} &
  \multicolumn{1}{c|}{{\begin{tabular}[c]{@{}c@{}}DL/\\ UL\end{tabular}}} &
  \multicolumn{1}{c|}{{\begin{tabular}[c]{@{}c@{}}DL/\\ UL\end{tabular}}} &
  \multicolumn{1}{c|}{{\begin{tabular}[c]{@{}c@{}}DL/\\ UL\end{tabular}}} &
  \multicolumn{1}{c|}{{DL}} & {UL} \\ \hline
\multicolumn{1}{c|}{\cellcolor[HTML]{EFEFEF}\textbf{\begin{tabular}[c]{@{}c@{}}HMD \\ Orientations\end{tabular}}} &
  \multicolumn{1}{c|}{UL} &
  \multicolumn{1}{c|}{UL} &
  \multicolumn{1}{c|}{UL} &
  \multicolumn{1}{c|}{UL} &
  \multicolumn{1}{c|}{UL} &
  DL \\ \hline
\multicolumn{1}{c|}{\cellcolor[HTML]{EFEFEF}\textbf{\begin{tabular}[c]{@{}c@{}}Tactile \\ Control\end{tabular}}} &
  \multicolumn{1}{c|}{UL} &
  \multicolumn{1}{c|}{UL} &
  \multicolumn{1}{c|}{UL} &
  \multicolumn{1}{c|}{UL} &
  \multicolumn{1}{c|}{UL} &
  DL \\ \hline
\multicolumn{1}{c|}{\cellcolor[HTML]{EFEFEF}\textbf{\begin{tabular}[c]{@{}c@{}}Haptic \\ Feedback\end{tabular}}} &
  \multicolumn{1}{c|}{DL} &
  \multicolumn{1}{c|}{DL} &
  \multicolumn{1}{c|}{DL} &
  \multicolumn{1}{c|}{DL} &
  \multicolumn{1}{c|}{DL} &
  UL \\ \hline
\multicolumn{1}{c|}{\cellcolor[HTML]{ECF4FF}\textbf{Environment}} &
  \multicolumn{1}{c|}{\begin{tabular}[c]{@{}c@{}}Active\\ Virtual\end{tabular}} &
  \multicolumn{1}{c|}{\begin{tabular}[c]{@{}c@{}}Passive\\ Virtual\\ and Real\end{tabular}} &
  \multicolumn{1}{c|}{\begin{tabular}[c]{@{}c@{}}Active\\ Virtual\\ and Real\end{tabular}} &
  \multicolumn{1}{c|}{\begin{tabular}[c]{@{}c@{}}Active\\ Virtual\end{tabular}} &
  \multicolumn{2}{c}{\begin{tabular}[c]{@{}c@{}}Active\\ Virtual\\ and Real\end{tabular}} \\ \hline
\multicolumn{1}{c|}{\cellcolor[HTML]{ECF4FF}\textbf{Display Device}} &
  \multicolumn{1}{c|}{\begin{tabular}[c]{@{}c@{}}HMD\\ {[}Occlusion{]}\end{tabular}} &
  \multicolumn{1}{c|}{\begin{tabular}[c]{@{}c@{}}OHMD\\ {[}Translucent{]}\end{tabular}} &
  \multicolumn{1}{c|}{\begin{tabular}[c]{@{}c@{}}OHMD\\ {[}Translucent{]}\end{tabular}} &
  \multicolumn{1}{c|}{\begin{tabular}[c]{@{}c@{}}HMD\\ {[}Occlusion{]}\end{tabular}} &
  \multicolumn{2}{c}{\begin{tabular}[c]{@{}c@{}}HMD\\ {[}Occlusion{]}\end{tabular}} \\ \hline
\multicolumn{1}{c|}{\cellcolor[HTML]{ECF4FF}\textbf{Refresh Rate}} &
  \multicolumn{1}{c|}{{[}90,120{]} Hz} &
  \multicolumn{1}{c|}{{[}60,90{]} Hz} &
  \multicolumn{1}{c|}{{[}90,120{]} Hz} &
  \multicolumn{1}{c|}{{[}90,120{]} Hz} &
  \multicolumn{2}{c}{{[}60,90{]} Hz} \\ \hline
\multicolumn{1}{c|}{\cellcolor[HTML]{ECF4FF}\textbf{Usage Scenario}} &
  \multicolumn{1}{c|}{In-premise} &
  \multicolumn{1}{c|}{\begin{tabular}[c]{@{}c@{}}In-premise/\\ Outdoor\end{tabular}} &
  \multicolumn{1}{c|}{\begin{tabular}[c]{@{}c@{}}In-premise/\\ Outdoor\end{tabular}} &
  \multicolumn{1}{c|}{In-premise} &
  \multicolumn{2}{c}{\begin{tabular}[c]{@{}c@{}}In-premise/\\ Outdoor\end{tabular}} \\ \hline
\end{tabular}
\end{table*}
\setlength{\textfloatsep}{1pt}

\subsection{Characteristics of XR Traffic Modalities} %
The traffic and network characteristics of different traffic modalities \cite{xr_std,xr_iotj}, as detailed in Table~\ref{table1}, are presented here.

\begin{itemize}
\item \textbf{Video:} 
Video is the dominant traffic component in each XR session and can be a single combined stream for both eyes or a multi-stream configuration with one stream per eye. According to the 3GPP XR traffic model \cite{xr_3gpp}, video frames are pseudo-periodically generated at the source, i.e., if $x(t)$ is a \emph{periodic signal} with period $T$, then, $x(t) = x(t + T), \forall t$, but a \emph{pseudo-periodic signal} $y(t)$ with average period $T$ can be defined as $y(t)\approx y(t+T+\epsilon(t)), \forall t$, where $\epsilon(t)$ is a slowly varying function following any arbitrary distribution. When XR video traffic is modelled as a time-series, no significant long-term trend is observed. Instead, \emph{seasonality} is represented by an average period of $T$ and with a \emph{jitter} of $\epsilon(t)$. Typically, the video frame generation period follows a Truncated Gaussian distribution with Mean = 1/(frame rate), standard deviation (SD) = 2 msec, and [Min, Max] = $[-4,4]$ msec or $[-5,5]$ msec. In addition, video frame sizes also follow a Truncated Gaussian distribution with Mean = (average datarate)/(frame rate $\times$ 8) bytes and [SD, Min, Max] = $[10.5\%, 50 \%, 150\%]$ of Mean. 
\item \textbf{HMD orientations:} In XR applications, the orientations and physical positions of an HMD/OHMD play a critical role, particularly for enabling optimized viewport-dependent streaming (VDS). Since VDS dynamically selects and transmits only the portion of the scene that the user is currently looking at or the ``viewport", its datarate is reduced significantly \cite{vr_mpg} as compared to that required for $360\degree$ video streaming. Motion in HMD or optical HMD (OHMD) is captured in two formats: three degrees-of-freedom (3DoF), which tracks orientation only, or six degrees-of-freedom (6DoF), which tracks both orientation and physical position, resulting in 100-500 bytes. Orientation can be represented as Euler angles (yaw, pitch, and roll), or as normalized quaternions to avoid issues like gimbal lock. Similarly, the physical position is tracked as $(x, y, z)$-space coordinates. Based on our experimental observations in \cite{xr_iotj}, the modelling of the pseudo-periodic HMD orientations traffic inter-arrival time in this work, follows a Gamma distribution with Mean = 15 msec and SD = 2 msec.
\item \textbf{Tactile control:} Human users manipulate virtual or remote real objects in an XR session using tactile controllers or gloves embedded with orientation and flex sensors. The spatial movements of the hands and fingers are tracked using 3DoF/6DoF representations. Additionally, tension and bending angles across various hand segments measured by the flex sensors are digitized and aggregated into packets of 1000-1500 bytes. In accordance with observed experimental data from commercially available devices in our testbed, tactile control traffic is modeled as pseudo-periodic with Mean = 10~msec and jitter as a generalized Pareto distribution with SD = 2~msec in this work \cite{h2m_sourav}.
\item \textbf{Haptic feedback:} Finally, haptic feedback generated upon impact between the remote machine and physical or virtual objects is used to replicate human touch and motion. This feedback encompasses both \emph{kinesthetic perception}, capturing forces, torques, positions, and velocities perceived through muscles, joints, and tendons, and \emph{tactile perception}, reflecting surface texture, friction, and pressure sensed via mechanoreceptors in the skin \cite{Polan_2017}. The generated data are aggregated into packets of 1000-1500 bytes. Consistent with experimental data obtained from commercial systems, this haptic traffic is modeled as pseudo-periodic with Mean = 10~msec and jitter following a generalized Pareto distribution with SD = 2~msec in this work \cite{h2m_sourav}.
    
\end{itemize}

\subsection{Transmission of XR applications} 
%
Here, the transmission characteristics of single-user immersive XR applications encompassing virtual reality (VR), augmented reality (AR), and mixed reality (MR), as well as their multi-user interactive extensions like human-to-human (H2H) and human-to-machine (H2M) collaborations, are discussed. 

\begin{itemize}
    \item \textbf{Virtual Reality (VR):} VR represents the top end of the immersion spectrum, completely replacing the physical environment with a computer-generated digital environment. In VR, human users interact with the digital environments through head-mounted displays (HMDs) that provide high-resolution stereoscopic imagery, spatial audio, motion tracking, and haptics. HMD orientations and tactile control signals are transmitted in the uplink (UL) direction to the application server located at the central office where  human presence in the virtual space and interactions with active virtual objects are enabled. In turn, viewport-dependent video and haptic feedback are transmitted in the downlink (DL) direction from the application server \cite{vr_bk}. Therefore, VR UL traffic (from humans) demands only low-latency and high-reliability, where as VR DL traffic (from application server) necessitates high-capacity, low-latency, and high-reliability.
    \item \textbf{Augmented Reality (AR).} AR is the most mature XR technology out of the three.  Through AR technology, digital content and passive virtual objects are overlaid onto the physical world in real-time, allowing human users to perceive both experiences simultaneously. In AR, human users use optical see-through or video see-through displays such as OHMDs, smartphones, or smart glasses to align virtual objects with the user’s real-world FOV. Computationally-intensive workloads are often offloaded to edge application servers through \emph{split-rendering}, allowing OHMDs and consumer AR devices to be more computationally lightweight \cite{moyukh}. In AR, high-quality video and HMD orientations are first transmitted UL to the application server. Then a relatively lower-quality video, strictly synchronized with the UL and with overlaid virtual objects are transmitted in the DL back to the user. Transmission of tactile control and haptic feedback are optional, depending on the type of usage scenario. Thus, high-throughput, low-latency, and high-reliability are required for both UL and DL traffic.
    \item \textbf{Mixed Reality (MR).} MR extends AR technology by enabling virtual and physical objects to interact within a shared spatial environment. With MR technology, virtual objects can be \emph{occluded}, i.e., hidden behind real-world objects and interact realistically with the geometry, physics, and lighting of the physical environment. The UL and DL transmissions in MR are similar to AR, but with more advanced spatial mapping, scene understanding, and latency-sensitive data synchronization required at the application server \cite{mr}.
    \item \textbf{Human-to-Human (H2H) collaborations.} In H2H collaboration, geographically distant human users can immersively communicate and collaborate within a shared virtual or mixed environment. Each user’s HMD orientations and tactile control signals are transmitted in the UL, requiring low-latency and high-reliability \cite{vr_mpg}. Consequently, high-quality viewpoint-dependent video, reconstructed avatars, and synchronized haptic feedback are transmitted DL such that perceptual consistency is maintained across all users.
    \item \textbf{Human-to-Machine (H2M) collaborations.} Unlike other XR applications, human users in H2M interact with virtual or remote machines, such as industrial robots or drones, through tactile and haptic interfaces. Each human user's HMD orientations and tactile control signals are transmitted UL to the application server located at the central office and then forwarded DL to the remote machine. Correspondingly, haptic feedback and real-time situational video with overlaid environmental data are transmitted UL from the remote machine to the application server and then delivered to the human in the DL. In H2M, UL and DL traffic thus form a closed-loop interactive control system that limits the actual deployable human-machine distance due to having to meet  meet QoE requirements (refer to Table~\ref{table1}).  Research in \cite{h2m_sourav} shows that the control loop can be decoupled from the actual human-machine distance, by preempting and subsequently transmitting the predicted haptic feedback to the user ahead of time \cite{h2m_sourav}.
\end{itemize}
%


\section{Generative-Predictive Resource Allocation} \label{sec4}

Proactive resource allocation has previously been investigated for latency-sensitive optical access networks and, more recently, immersive H2M/XR applications \cite{Polan_2017, WongJOCNTutorial2023, xr_iotj}. In \cite{xr_iotj}, we proposed an efficient end-to-end resource allocation scheme based on user head-movement predictions to enable XR collaborations between remote industrial premises. However, such user-sensitive application data is often unavailable for collection from real networks. Hence, employing traffic prediction models based on historical traffic traces is essential in designing proactive resource allocation frameworks. In \cite{cnn_lstm}, high traffic prediction accuracy on open-source Zenodo and PeMS datasets was achieved using convolutional neural networks (CNNs), recurrent neural networks (RNNs), LSTM networks, and their hybrid combinations. Further, in \cite{trans_gen}, a generative AI-based traffic forecasting model was proposed through integrating a graph neural network with a Transformer. Focusing on traffic in optical metro and access networks, a novel spatio-temporal Transformer was proposed in \cite{jlt_gen_ai} that integrates graph attention mechanisms to capture spatial correlations and a Transformer encoder to effectively model long-range temporal dependencies. In addition, a traffic prediction model integrating a Crested Porcupine optimizer, bi-directional-TCN, and BiLSTM networks was presented in \cite{fttr_genai} achieving accurate  prediction of complex FTTR traffic. In \cite{TinyML2026}, lightweight machine learning models, i.e. TinyML, in combination with predictive resource allocation were proposed to support latency-sensitive tactile control and haptic feedback traffic in resource-constrained FTTH networks.

While these foundational studies show great promise in making highly-accurate traffic predictions, their performance on multi-tier aggregated multi-modal XR traffic over the integrated FTTR architecture warrants thorough investigation. Here, we propose a Generative-Prediction (Gen-Pred) AI framework with highly-accurate predictive and generative time-series forecasting methods that are embedded with coarse-grained proactive resource allocation schemes for internal FTTR and external FTTH segments. Results from a preliminary investigation was published in \cite{fttr_ofc_26}, with this article presenting a significant
extension of resource allocation across the FTTH+FTTR+WiFi network with network-orchestrated roaming.

\subsection{System Model and Coordinated Resource Allocation} \label{sec4.4}
\begin{figure*}[!t]
\centering
\includegraphics[width=0.9\textwidth,keepaspectratio]{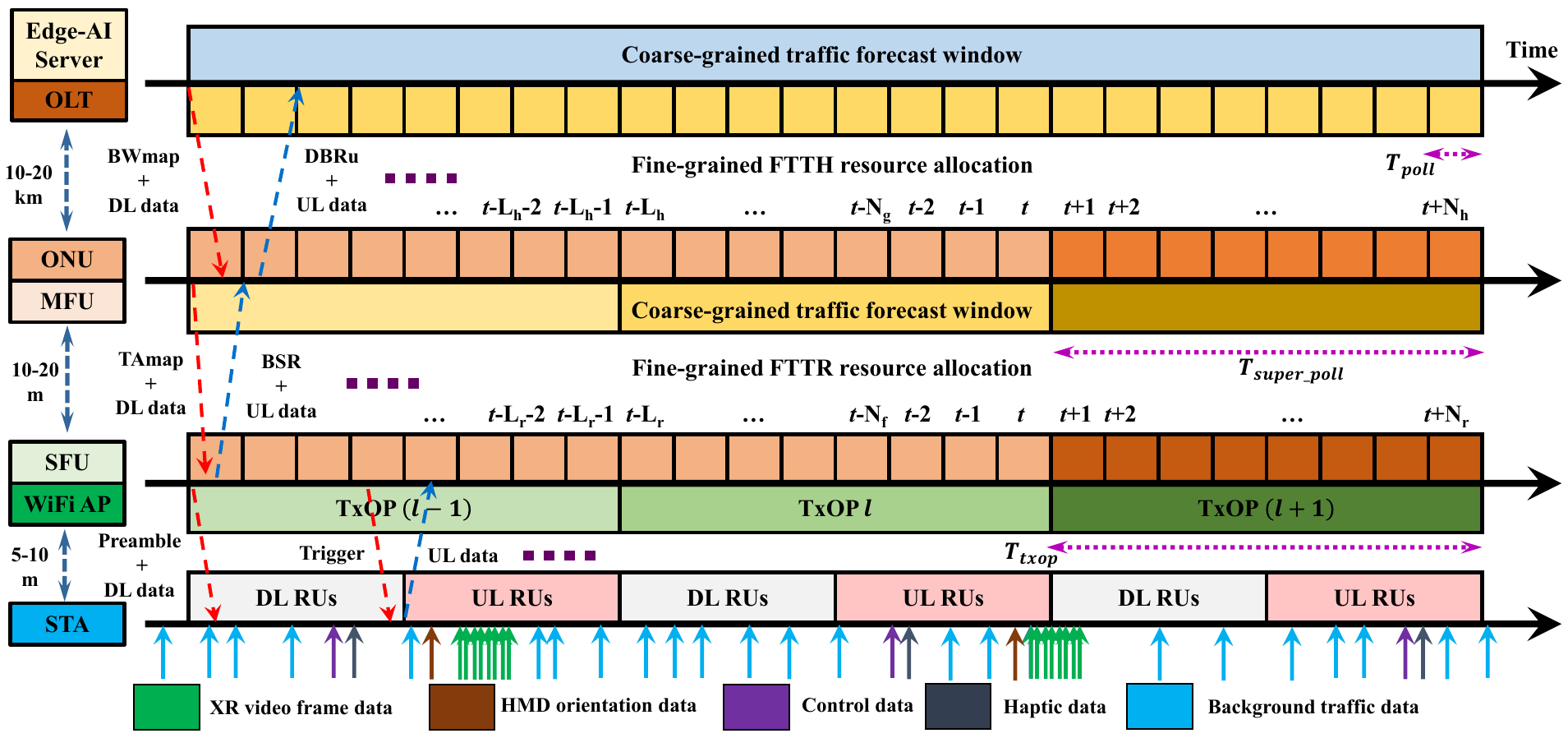}
\caption{Timing diagram of multi-tier coordinated proactive resource allocation for an integrated FTTR network architecture. Note that time durations of different intervals shown are only for illustration and are not to the scale.}
\label{timing}
\end{figure*}

In this work, an FTTR network using XG(S)-PON technology supporting WiFi-7 WAPs in each room of the premise is considered. Each WAP has 16 antennas, all equally divided for simultaneous operations in both 5~GHz and 6~GHz bands. Fig.~\ref{timing} illustrates the timing diagram of a multi-tier coordinated proactive resource allocation scheme for the integrated FTTH + FTTR + WiFi network considered in our work.  All WAPs in a premise are configured with the same TxOP duration $T_{txop}$ (maximum values for voice and video access classes are 1.505 and 3.008 msec, respectively), durations much longer than $T_{poll}$ = 125~$\mu$sec polling cycle of XGS-PON. Further, the parameter $T_{txop}$ is an \emph{integer multiple} of $T_{poll}$, e.g., 1-3 msec to achieve times-slotted FTTR-WiFi coordinated resource allocation scheme.

Referring to Fig.~\ref{timing}, after an MFU receives BSRs and uplink reports from the associated SFU-WAPs in a TxOP, it schedules the RUs and transmission times of the WAPs in the DL and the STAs in the UL for the next TxOP. Recall that the OMCI-based interface at the MFU has inherent mechanisms for time synchronization among all network interfaces.
In addition, the corresponding TAmaps for the intermediate $\lceil T_{txop}/T_{poll} \rceil$ polling cycles are also generated. Note that the G.fin TAmap is functionally analogous to the G-PON BWmap, but its duration needs not to be fixed at 125 $\mu$sec \cite{ITUTG9942}. It can operate on a longer, configurable \emph{super-frame}, i.e., $T_{super\_poll}$ = 1-3 msec to align with WiFi timeslots. The complex aggregated multi-modal XR and background traffic at the residential gateways are reported further in each polling cycle to the OLT of the external FTTH network based on 50G-PON. Accordingly, BWmaps are generated for transmission of this data in the UL. 

Since the FTTH, FTTR, and WiFi modules largely operate independently in practical scenarios, a \emph{coarse-grained resource allocation scheme} can be adopted for its efficiency and ease of rapid deployment \cite{7333289}. Nonetheless, its reactive nature to bandwidth requests may not address the stringent QoE requirements of XR collaborations \cite{lstm_opt}. Hence, a proactive resource allocation framework harnessing state-of-the-art generative and predictive long-horizon multi-step-ahead traffic forecasting methods is proposed. The framework is detailed in the subsequent subsections.\par

\setlength{\textfloatsep}{5pt}


\subsection{Predictive Traffic Forecasting in FTTR Segment} 
Recall that $T_{txop}$ denotes the TxOP duration configured for all the WAPs in a premise and $T_{poll}$ denotes the polling cycle duration. Therefore, within each TxOP, there are $N_r = \lceil T_{txop}/T_{poll}\rceil$ XG(S)-PON polling cycles or timeslots. Let $\mathcal{I} = \{1,\dots,W\}$ denote the set of SFU/WAPs, $\mathcal{J} = \{1,\dots,S_{xr},S_{xr}+1,\dots,S_{xr}+S_{bkg}\}$ denote the set of STAs connected to each WAP, where $S_{xr}$ indicates the number of XR devices with either video, HMD orientation, control, or haptic traffic and $S_{bkg}$ indicates the number of background devices, $x_t^{(i,j)}$ denote the UL traffic from STA-$j$ to WAP-$i$ in timeslot-$t$ and $X_l^{(i,j)} = \sum_{k=1}^{N_r} x_{t+k}^{(i,j)}$ denotes the total traffic arrival in TxOP-$l$ enveloping $N_r$ timeslots. The input sequence of aggregated UL traffic at WAP-$i$ over the last $L_r$ (usually $L_r\gg N_r$) timeslots is represented as:\\
\vspace{-\baselineskip}
\begin{align}
 \bm{X}_{t-1}^{(i)} &= [\bm{x}_{t-L_r}^{(i)}, \dots, \bm{x}_{t-1}^{(i)}], \forall i\in\mathcal{I}, \label{eq1}\\
 \bm{x}_{t-1}^{(i)} &= [x_{t-1}^{(i,j)}]_{j=1}^{S} =[x_{t-1}^{(i,1)}, \dots, x_{t-1}^{(i,S)}]^\top, \forall i\in\mathcal{I}. \label{eq2}
\end{align}
\par Considering that traffic characteristics of each STA will be independent of each other, a separate long-horizon multi-step-ahead traffic forecast model $f_{\Theta}^{(i,j)}(\cdot)$ for each STA is created at the MFU. Common forecast models can similarly be created for STAs with similar traffic whereby their historical data can be combined for model training. 

If the model inference latency $\tau_r \geq T_{poll}$, then the inferencing must be initiated $N_f = \lceil{\tau_r/T_{poll}}\rceil$ timeslots before the current TxOP ends where $\tau_r$ is the 95$^{\text{th}}$ percentile of the inference time distribution. Subsequently, denoting the inference completion time to be timeslot-$t$, the inference process would be initiated at timeslot $(t-N_f)$. Using an input sequence ending at $(t-N_f)$, denoted as $\mathbf{x}_{t-N_f}^{(i,j)} = [x_{t-N_f-L_r+1}^{(i,j)}, \dots, x_{t-N_f}^{(i,j)}]^\top$, the model generates forecasts for the next $(N_f+N_r)$ timeslots:\\
\vspace{-\baselineskip}
\begin{align}
\hat{\bm{y}}_{t-N_f}^{(i,j)} &= f_{\Theta}^{(i,j)}(\mathbf{x}^{(i,j)}_{t-N_f}), \nonumber\\
&= [\hat{x}_{t-N_f+1}^{(i,j)}, \dots, \hat{x}_t^{(i,j)}, \hat{x}_{t+1}^{(i,j)}, \dots, \hat{x}_{t+N_r}^{(i,j)}]^\top. \label{{eq3}}
\end{align}
Since $N_f$ timeslots are consumed to make the traffic forecasts, the predictions $[\hat{x}_{t-N_f+1}^{(i,j)}, \dots, \hat{x}_t^{(i,j)}]$ correspond to the past and are subsequently discarded. Only the sequence $[\hat{x}_{t+1}^{(i,j)}, \dots, \hat{x}_{t+N_r}^{(i,j)}]^\top$ is utilized to generate TAmaps for the next $N_r$ polling cycles. In each polling cycle or timeslot-$t$, after receiving BSR $r_t^{(i,j)}$ and data $\alpha_t^{(i,j)}$, the MFU calculates the actual traffic arrival for the previous slot as $x_{t-1}^{(i,j)} = \max\{0, r_t^{(i,j)}+\alpha_t^{(i,j)}-r_{t-1}^{(i,j)}\}$. This relation holds true for any PON where the one-way transmission latency
is less than $T_{poll}$. 

For the immediate next timeslot-$(t+1)$, the MFU grants each SFU-$i$ with:\\
\vspace{-\baselineskip}
\begin{align}
g_{t+1}^{(i,j)} = \min\left\{\rho_{t+1}^{(i,j)}, \left(\frac{\rho_{t+1}^{(i,j)}}{\sum_k \rho_{t+1}^{(i,k)}}\right) \cdot BW_{max}^{(i)}\right\},
\label{eq4}
\end{align}
where $\rho_{t+1}^{(i,j)} = (r_t^{(i,j)} + \hat{x}^{(i,j)}_{t+1})$, representing the \emph{projected buffer occupancy}, i.e., the sum of the currently reported buffer status ($r_t$) and the new traffic predicted to arrive during the grant execution ($\hat{x}_{t+1}$), and $BW_{max}^{(i)} = (R_r \times T_{poll}/W)$ denotes the maximum bandwidth available per polling cycle where $R_r$ denotes the line rate of the PON in FTTR segment. Hence, this resource allocation scheme is robust against under and over-predictions. To train this model with a historical dataset, the mean-squared error (MSE) loss across the $(N_f+N_r)$ timeslots per-sample-$\mu$ is given by:\\
\vspace{-\baselineskip}
\begin{align}
\mathcal{L}_{\mu}^{(i,j)} = \frac{1}{N_f+N_r}\sum_{k=-N_f+1}^{N_r} \left(x_{t+k}^{(i,j)} - \hat{x}_{t+k}^{(i,j)}\right)^2. \label{eq5}
\end{align}



\subsection{Generative Traffic Forecasting in FTTH Network} \label{sec5.2}
Traffic generated within each premise (FTTH-WiFi) is highly complex and is expected to be aggregated at its corresponding ONU of the external FTTH network prior to uplink transmission to the OLT. Consequently, traditional resource allocation schemes in external FTTH networks face severe operational bottlenecks in satisfying the QoE requirements of XR collaborations. Recall that we consider that all traffic belong to TC-4 flows and we may proactively allocate resources only for multi-modal XR traffic in both FTTH and FTTR-WiFi segments as background traffic is low-priority and latency-tolerant.
Hence, let $\xi_t^{(u)}$ and $\beta_t^{(u)}$ denote the aggregated multi-modal XR traffic and background traffic arrival at ONU-$u$ in polling cycle or timeslot-$t$, respectively. The coarse-grained interval for traffic forecasting $T_{int}$ is chosen to be a large-integer multiple of polling cycle duration such that $N_h = \lceil T_{int}/T_{poll}\rceil$, as shown in Fig.~\ref{timing}. If the traffic forecast model inference latency (95$^{\text{th}}$ percentile) $\tau_h \geq T_{poll}$, then the inferencing should be started $N_g = \lceil{\tau_h/T_{poll}}\rceil$ timeslots before the target interval $T_{int}$ begins. This requires forecasting for an additional $N_g$ timeslots to cover the inference latency. This expanded inference window inherently requires the adoption of an advanced Transformer-based model, which significantly outperforms conventional recurrent and convolutional neural networks in long-horizon, multi-sample-ahead forecasting \cite{bishop2023deep}.\par

Now consider the sequence of aggregated multi-modal XR traffic from all the $U$ ONUs over the last $L_h$ (usually $L_h\gg N_h$) polling cycles, available at the start of inference as:\\
\vspace{-\baselineskip}
\begin{align}
\mathbf{\Xi}_{t-N_g} &= [\bm{\xi}_{t-N_g-L_h+1}, \ldots, \bm{\xi}_{t-N_g}], \label{eqf1}\\
\bm{\xi}_{k} &= [\xi_{k}^{(1)}, \ldots, \xi_{k}^{(U)}]^\top. \label{eqf2}
\end{align}
Using \eqref{eqf1} as input data, for each ONU-$u$ the Transformer-based traffic forecasting module implemented in the Edge-AI server generates $\hat{\bm{\Xi}}^{(u)}_{t-N_g+1:t+N_h} = \gamma^{(u)}_\Phi(\mathbf{\Xi}^{(u)}_{t-N_g})$, where $\gamma^{(u)}_{\Phi}(.)$ is a parameterized multi-step predictor implemented using \emph{attention mechanisms}. The complete forecast generated is: 

\vspace{-\baselineskip}
\begin{align}
\hat{\bm{\Xi}}_{t-N_g+1:t+N_h} = [\hat{\bm{\xi}}_{t-N_g+1}, \dots, \hat{\bm{\xi}}_{t}, \hat{\bm{\xi}}_{t+1}, \dots, \hat{\bm{\xi}}_{t+N_h}].
\label{eq8}
\end{align}
Upon completion of inference which commenced and ended at timeslot-$(t-N_g)$ and timeslot-$t$, respectively, the first $N_g$ predictions $[\hat{\bm{\xi}}_{t-N_g+1}, \dots, \hat{\bm{\xi}}_{t}]$ are discarded as these timeslots have already passed. The remaining traffic forecasts $\hat{\bm{\xi}}^{(u)}_{t+k}, \forall k\in\{1,\dots,{N_h}\}$ for each ONU are used by the BWmap generator to proactively allocate resources for uplink transmission.

The maximum bandwidth that can be allocated to each ONU-$u$ in a polling cycle is given by $BW^{(u)}_{max} = (R_h\times T_{poll}/U)$, where $R_h$ denotes the line rate of the PON in the FTTH segment. This bandwidth is proportionally divided between the XR and background traffic to minimize bandwidth wastage. Therefore, after receiving a dynamic buffer report (DBRu) containing $rx_t^{(u)}$ for XR traffic and $rb_t^{(u)}$ for background traffic in timeslot-$t$, the corresponding grant provided to ONU-$u$ consists of $g{\xi}_{t+1}^{(u)} = \min\left\{ \rho \xi_t^{(u)},\frac{\rho \xi_t^{(u)}}{\rho\xi_t^{(u)}+rb_t^{(u)}}\cdot BW^{(u)}_{max}\right\}$ and $g{\beta}_{t+1}^{(u)} = \min\left\{rb_t^{(u)},\frac{rb_t^{(u)}}{\rho\xi_t^{(u)}+rb_t^{(u)}}\cdot BW^{(u)}_{max}\right\}$, where $\rho\xi_t^{(u)}= (rx_t^{(u)}+\hat{\xi}_{t+1}^{(u)})$. Note that $\hat{\xi}_{t+1}^{(u)}$ is used to account for arrivals during the grant execution. After receiving XR data $\alpha x_t^{(u)}$, the actual XR traffic arrival in the previous timeslot-$(t-1)$ is calculated as $\xi_{t-1}^{(u)} = \max\{0, rx_t^{(u)}+\alpha x_t^{(u)}-rx_{t-1}^{(u)}\}$. The training process of the traffic forecast models for each ONU-$u$ aims to minimize the multi-step MSE loss across the $(N_g+N_h)$ timeslots per-sample-$\nu$, given by:\\
\vspace{-\baselineskip}
\begin{align}
\mathcal{L}_{\nu}^{u} = \frac{1}{N_g+N_h}\sum_{k=-N_g+1}^{N_h}
\left(\xi_{t+k}^{(u)} - \hat{\xi}_{t+k}^{(u)}\right)^2. \label{eqf3}
\end{align}
%

\subsection{Intelligent In-Premise Seamless Roaming}\label{sec5.1}
Here, we present a seamless in-premise intelligent roaming scheme designed to meet the stringent QoE requirements of XR collaborations. A mixed-integer non-linear program (MINLP), $\mathcal{P}_{FTTR}$, is formulated for joint in-premise WAP-STA association, resource allocation, and seamless handover, with the objective of \emph{maximizing the total network utility in each TxOP}, i.e., satisfying the UL and DL demands of the maximum number of STAs. STAs report the channel conditions of their neighboring WAPs and are assigned to a WAP with good channel quality and sufficient capacity. For stability, an associated STA remains attached to its current WAP unless its QoE requirements (throughput, latency, reliability) are violated over $N_{\delta}$ consecutive TxOPs. Upon such violation, the MFU initiates a ``make-before-break" handover \cite{fttrb_ofc}: the affected STAs immediately report the channel conditions of other neighboring WAPs, the MFU solves $\mathcal{P}_{FTTR}$ in the subsequent TxOP to identify the optimal target WAP, and signals the OLT over the FTTH segment to reconfigure the network path before the STAs are switched over seamlessly (refer to Fig. \ref{arch1}). This minimizes frequent handovers of mobile STAs within the dense in-premise FTTR-WiFi segment.\par
\textbf{Sets and Parameters:} Recall that $\mathcal{I} = \{1,\dots,W\}$ denotes the set of WAPs, $\mathcal{J} = \{1,\dots,S_{xr},S_{xr}+1,\dots,S_{xr}+S_{bkg}\}$ the set of STAs, and $X_l^{(i,j)}$ the UL traffic arriving at WAP-$i$ from STA-$j$ during OFDMA TxOP-$l$. The supported MLO bands are $\mathcal{B} = \{2.4, 5, 6\}$ GHz, with available RU bandwidths (in MHz) $\mathcal{W}_{2.4} = \{2,4,8,20,40\}$, $\mathcal{W}_{5} = \{2,4,8,20,40,80,160\}$, and $\mathcal{W}_{6} = \{2,4,8,20,40,80,160,320\}$, and $BW_b^{max}$ denotes the total bandwidth capacity of band $b \in \mathcal{B}$. Channel quality is characterized by the spectral efficiency $E_{ijb}$ (bits/s/Hz) of the WAP-$i$--STA-$j$ link on band $b$, weighted by a priority factor $p_{ijb}$, while $R_{req,j}^{UL}$ and $R_{req,j}^{DL}$ denote the minimum data rate per TxOP for STA-$j$ (see Table \ref{table1}). Roaming hysteresis is enforced through $\theta_{ij} \in \{0,1\}$, indicating the association status in the previous TxOP. Since traffic forecasting and decision-making require $N_w$ TxOPs, the forecast horizon is set to $(N_w+1)$. At the end of TxOP-$l$, upon receiving the BSR $r_l^{(i,j)}$ and data $X_l^{(i,j)}$, the MFU computes the traffic generated in the previous TxOP as $\Gamma_{l-1}^{(i,j)} = \max\{0, r_l^{(i,j)}+X_l^{(i,j)}-r_{l-1}^{(i,j)}\}$, and from the input sequence $\mathbf{\Gamma}_{l-N_w}^{(i,j)} = [\Gamma_{l-N_w-L_w+1}^{(i,j)}, \dots, \Gamma_{l-N_w}^{(i,j)}]^\top$ of length $L_w$ generates the forecasts $\hat{\bm{\Gamma}}_{l-N_w+1:1}^{(i,j)}=[\hat{\Gamma}^{(i,j)}_{l-N_w+1},\dots,\hat{\Gamma}^{(i,j)}_{l},\hat{\Gamma}^{(i,j)}_{l+1}]$ via the predictive model of Sec. \ref{sec5.1}. As $N_w$ TxOPs elapse during inference, $\hat{\Gamma}^{(i,j)}_{l+1}$ is used to project the UL buffer occupancy as $Q_{i,j}^{UL} (l+1) = (r_l^{(i,j)}+\hat{\Gamma}^{(i,j)}_{l+1})$, whereas $Q_{i,j}^{DL} (l+1)$ is deterministically known by the MFU. These values represent the \emph{resource demands} of STA-$j$ and are transmitted to WAP-$i$ immediately.\par
\textbf{Decision variables:} $\phi_{ij} \in \{0,1\}$ equals 1 if STA-$j$ associates with WAP-$i$; $\mu_{ijbw}^{DL}, \mu_{ijbw}^{UL} \in \{0,1\}$ equal 1 if STA-$j$ is assigned a discrete RU of bandwidth $w \in \mathcal{W}_b$ on band $b$ of WAP-$i$; and $\tau_i \in [\tau_{min}, \tau_{max}]$ is the continuous fraction of the TxOP duration $T_{txop}$ allocated to downlink transmission at WAP-$i$, with a guard time $\tau_g$ between the DL and UL intervals.\par
\textbf{MINLP formulation:} The \emph{objective function} of $\mathcal{P}_{FTTR}$ maximizes the total utility of satisfied demands:\\
\vspace{-\baselineskip}
\begin{align}
\max_{\bm{\phi},\bm{\mu},\bm{\tau}} \sum_{i \in \mathcal{I}} \sum_{j \in \mathcal{J}} \phi_{ij} \Bigg[ & \sum_{b\in\mathcal{B}} p_{ijb} \Biggl( \min\left\{Q_{ij}^{DL}, R_{ij}^{DL}\right\} \Biggr. \nonumber\\
& \quad \Biggl. + \min\left\{Q_{ij}^{UL}, R_{ij}^{UL} \right\}\Biggr) \Bigg], \label{obj}
\end{align}
where the $\min\{\cdot\}$ caps the utility by the buffer occupancy when $R_{ij} > Q_{ij}$; conversely, if $R_{ij} < Q_{ij}$, the utility is limited by the feasible transmission and the excess data is queued for the next TxOP, increasing latency. The feasible DL and UL transmissions $\forall i\in\mathcal{I}, j\in \mathcal{J}$ are:\\
\vspace{-\baselineskip}
\begin{align}
R_{ij}^{DL} &= \tau_i \cdot T_{txop} \cdot \sum_{b\in\mathcal{B}} \bigg(E^{DL}_{ijb} \cdot \sum_{w\in\mathcal{W}_b} \mu^{DL}_{ijbw} \cdot w \bigg), \label{eqR1}\\
R_{ij}^{UL} &= (1-\tau_i-\tau_g) \cdot T_{txop} \cdot \sum_{b\in\mathcal{B}} \bigg(E_{ijb}^{UL} \cdot \sum_{w\in\mathcal{W}_b} \mu^{UL}_{ijbw} \cdot w \bigg), \label{eqR2}
\end{align}
with spectral efficiencies:\\
\vspace{-\baselineskip}
\begin{align}
E^{DL}_{ijb} &= \log_2 \Bigg(1+\frac{P_{Tx}^{i}\cdot G^{DL}_{ijb}}{N_0 + I_{ijb}^{DL}}\Bigg), \quad \forall i, j, b,\label{eqE1} \\
E^{UL}_{ijb} &= \log_2 \Bigg(1+\frac{P_{Tx}^{j}\cdot G^{UL}_{ijb}}{N_0 + I_{ijb}^{UL}}\Bigg), \quad \forall i, j, b, \label{eqE2}
\end{align}
where $P_{Tx}^{i}$ and $P_{Tx}^{j}$ are the transmit powers of the WAP and STA, $G_{ijb}$ the channel gain (path loss and fading), $I_{ijb}$ the measured interference, and $N_0$ the noise power.\par
\emph{Throughput Constraints:} For an associated STA ($\phi_{ij}=1$), the aggregated throughput per TxOP needs to meet minimum QoE requirements:\\
\vspace{-\baselineskip}
\begin{align}
R_{ij}^{DL} &\ge \phi_{ij} \cdot R_{req, j}^{DL}, \quad \forall i,j, \label{eqass1}\\
R_{ij}^{UL} &\ge \phi_{ij} \cdot R_{req, j}^{UL}, \quad \forall i,j, \label{eqass2}
\end{align}
which are trivially satisfied when $\phi_{ij}=0$.\par
\emph{Connectivity constraints:} Each STA associates with at most one WAP. Existing connections are maintained to avoid ping-pong handovers, with $\theta_{ij}$ (initialized to 0 and set to 1 connection establishment) captures the previous association state:\\
\vspace{-\baselineskip}
\begin{align}
    \sum_{i \in \mathcal{I}} \phi_{ij} &\le 1, \quad \forall j, \label{eqass3}\\
    \phi_{ij} &\ge \theta_{ij}, \quad \forall i, j. \label{eq:hysteresis}
\end{align}
Constraint \eqref{eq:hysteresis} is relaxed for STA-$j$ if QoE degradation persists over $N_{\delta}$ consecutive TxOPs, allowing a network-initiated handover to a new WAP.

\begin{algorithm}[t!]
\caption{FTTR-WiFi Resource Allocation and Roaming} \label{alg1}
\hspace*{\algorithmicindent} \textbf{Input:} $\mathcal{I}, \mathcal{J}, \mathcal{B}, \mathcal{W}_b, Q_{ij}^{DL/UL}, E_{ijb}^{DL/UL}, \theta_{ij}, p_{ijb,} N_{max}, \epsilon$\\
\hspace*{\algorithmicindent} \textbf{Output:} Optimal solution: $\bm{\phi}^*, \bm{\mu}^*, \bm{\tau}^*$\\
\hspace*{\algorithmicindent} \textbf{Initialize:} $n \leftarrow 0$, $\tau_i^{(0)} \leftarrow 0.5, \ \forall i$, Objective $U^{(0)} \leftarrow 0$;
\begin{algorithmic}[1]
\While{$n < N_{max}$}
    \State \texttt{\# STA Association and RU Allocation}
    \State Fix time allocation $\bm{\tau}^{(n)}$;
    \For{$i \leftarrow 1$ \textbf{to} $|\mathcal{I}|$} \Comment{\textit{Parallelizable across WAPs}}
        \State Update $R_{ij}^{DL/UL}$ capacities based on fixed $\tau_i^{(n)}$ using (\ref{eqR1})-(\ref{eqR2});
        \State Employ greedy matching to find $\bm{\phi}_i^{(n+1)}, \bm{\mu}_i^{(n+1)}$;
        \State Ensure constraints (\ref{eqass1})-(\ref{eqcap3}) are satisfied;
    \EndFor
    \State \texttt{\# Time Allocation Optimization}
    \State Fix association $\bm{\phi}^{(n+1)}$ and RU allocation $\bm{\mu}^{(n+1)}$;
    \For{$i \leftarrow 1$ \textbf{to} $|\mathcal{I}|$}
        \State Calculate optimal $\tau_i^{(n+1)}$ via gradient update;
        \State Clip $\tau_i^{(n+1)}$ to $[\tau_{min}, \tau_{max}]$ to satisfy (\ref{eqtime1});
    \EndFor
    \State \texttt{\# Convergence Check}
    \State Calculate current total objective $U^{(n+1)}$ using (\ref{obj});
    \If{$|U^{(n+1)} - U^{(n)}| \le \epsilon$}
        \State \textbf{break}; \Comment{\textit{convergence or fixed point}}
    \Else
        \State $n \leftarrow n + 1$;
    \EndIf
\EndWhile
\State \textbf{return} $\bm{\phi}^{(n+1)}, \bm{\mu}^{(n+1)}, \bm{\tau}^{(n+1)}$;
\end{algorithmic}
\end{algorithm}
\setlength{\textfloatsep}{5pt}
%
\emph{OFDMA-MLO RU allocation constraints:} Each STA is assigned at most one discrete RU per band, and the cumulative allocated bandwidth per band cannot exceed $BW_b^{max}$:\\
\vspace{-\baselineskip}
\begin{align}
\sum_{w \in \mathcal{W}_b} \mu^{DL}_{ijbw} \le \phi_{ij} \quad \text{and} \quad &\sum_{w \in \mathcal{W}_b} \mu^{UL}_{ijbw} \le \phi_{ij}, \forall i,j,b, \label{eqcap1}\\
\sum_{j\in\mathcal{J}} \sum_{w \in \mathcal{W}_b} \mu^{DL}_{ijbw} \cdot w &\le BW_b^{max}, \quad \forall i,\label{eqcap2}\\
\sum_{j\in\mathcal{J}} \sum_{w \in \mathcal{W}_b} \mu^{UL}_{ijbw} \cdot w &\le BW_b^{max}, \quad \forall i.\label{eqcap3}
\end{align}
\par \emph{Time allocation constraints:} To ensure valid time-splits and reserve capacity for control overhead, e.g., beacons, $\tau_i$ is bounded by guard times \cite{wifi6_ofdma}:\\
\vspace{-\baselineskip}
\begin{align}
\tau_{min} \le \tau_i &\le \tau_{max}, \quad \forall i, \label{eqtime1}
\end{align}
while the normalization $\tau_i + (1-\tau_i-\tau_g) + \tau_g = 1$ is implicitly satisfied by \eqref{eqR1}-\eqref{eqR2}.\par
\textbf{Solution algorithm:} As $\mathcal{P}_{FTTR}$ is equivalent to the NP-hard \emph{generalized assignment problem} and couples binary ($\phi_{ij}, \mu_{ijbw}$) with continuous ($\tau_i$) variables, we adopt a \emph{decomposition-based alternating optimization approach} for its robustness and time-efficiency \cite{LiSun_Chapter13}, splitting the problem into two tractable sub-problems solved iteratively (Alg.~\ref{alg1}). Solved via a greedy heuristic as in \cite{minlp}, the association and RU allocation phase costs $\mathcal{O}(|\mathcal{I}| \cdot |\mathcal{B}| \cdot \max_b\{|\mathcal{W}_b|\} \cdot |\mathcal{J}|)$, dominated by demand sorting and RU assignment, while the time allocation phase is linear over $|\mathcal{I}|$ variables, requiring $\mathcal{O}(|\mathcal{I}|^{3.5})$ with interior point methods or $\mathcal{O}(|\mathcal{I}|)$ with gradient updates. The total complexity is therefore $\mathcal{O}\left(N_{iter} \cdot (|\mathcal{I}| \cdot |\mathcal{J}| \cdot |\mathcal{B}| \cdot \max_b\{|\mathcal{W}_b|\} + |\mathcal{I}|^{3.5})\right)$ for $N_{iter}$ convergence iterations. Since $|\mathcal{I}|$, $|\mathcal{J}|$, and $\max_b\{|\mathcal{W}_b|\}$ are small in in-premise FTTR-WiFi scenarios, the algorithm is polynomial-time and practical for deployment.

\section{Performance Evaluation} \label{sec6}

\subsection{Network Simulation Configurations} \label{sec6.1}
\begin{figure*}[!t]
\centering
\includegraphics[width=0.6\textwidth,keepaspectratio]{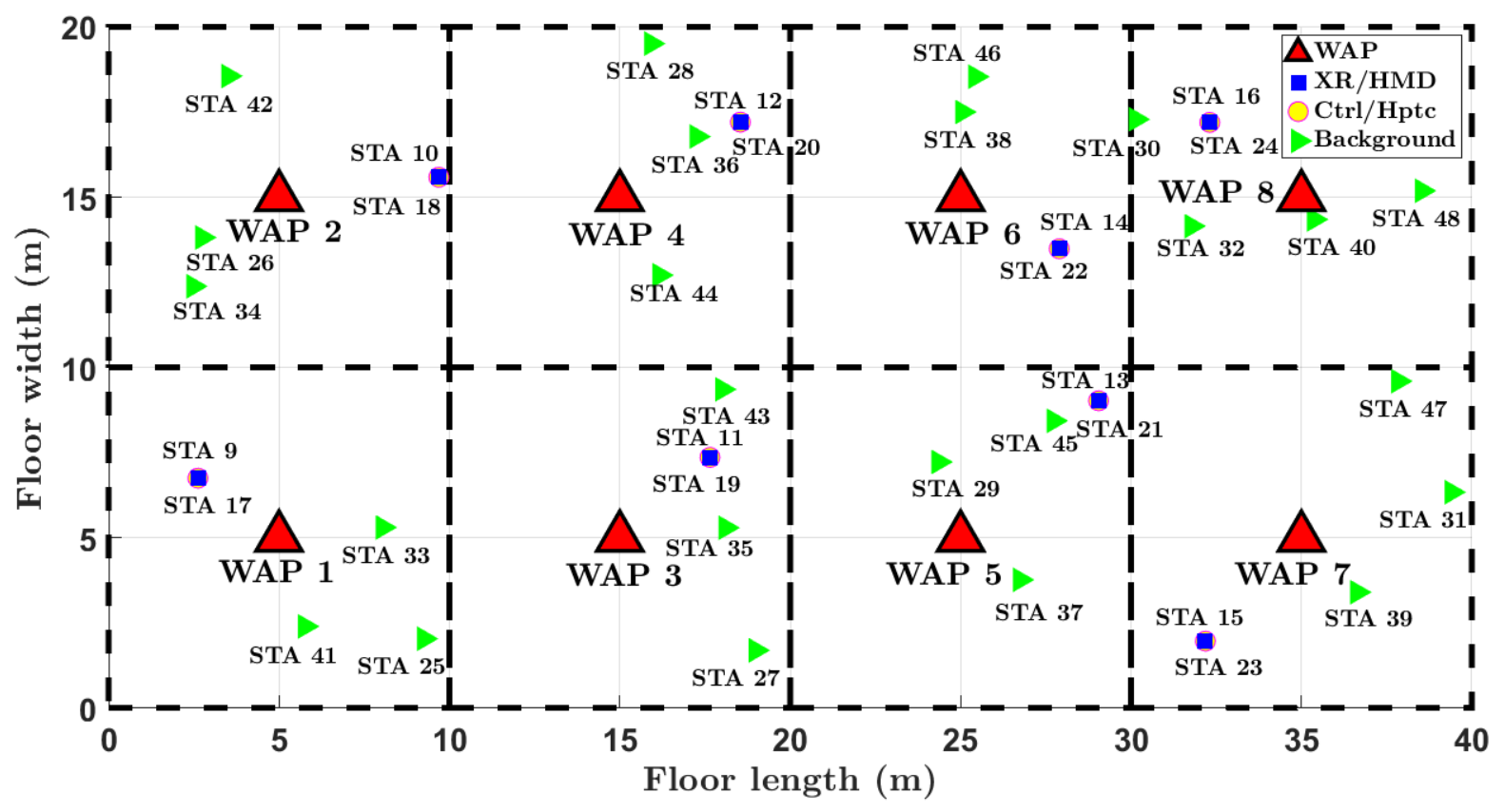}
\caption{In-premise layout with 8 integrated SFU-WAPs (rooms), each connected to 5 STAs supporting XR collaborations and background traffic.}
\label{exp_scn}
\end{figure*}

The resource allocation schemes of all the segments of the integrated FTTH-FTTR-WiFi architecture is implemented using the discrete event-driven simulator OMNeT++. Following the ITU-T recommendations \cite{itu_fttr}, we consider a 50G-PON of 20 km length as the first-segment FTTH network with the maximum available datarate in both the UL and DL are 50 Gbps. The duration of the polling cycles is fixed at 125~$\mu$sec and the guard time between GEM packets from consecutive ONUs is 0.25~$\mu$sec. The number of ONUs controlled by the OLT of the 50G-PON is 32 (split-ratio 1:32). Each of these ONUs are connected to an MFU which control a FTTR segment with XGS-EPON with maximum UL and DL datarate 10~Gbps, split ratio 1:8, MFU-SFU distance 20 m, polling cycle duration 125~$\mu$sec, and the guard time between packets from consecutive SFUs is 0.25~$\mu$sec. Hence, there are a total of 32 × 8 = 256 SFUs. Each SFU is further connected to a WiFi-7 WAP configured with OFDMA fixed TxOP duration of 1 msec and MLO with 8 antennas on 5 GHz and 8 antennas on 6 GHz bands with transmit 20 dBm power in the DL, as shown in Fig. \ref{exp_scn}. Moreover, we consider that four H2M  pairs in each premises are participating in an XR collaboration. In Fig. \ref{exp_scn}, each room has a dimension of 10$\times$10 m$^2$, in which the blue square indicates either a human (UL: HMD orientations, DL: XR video) or a machine (UL: XR video, DL: HMD orientations). Correspondingly, the yellow circle beneath it indicates a control device/gloves for humans (UL: control, DL: haptics) or a haptic device (UL: haptics, DL: control). In addition, there are 3 background devices (mobiles, laptops etc.) transmitting exponentially distributed traffic both in the UL and DL. Thus, there are 5 wireless devices or STAs in each room and the locations of all the 8$\times$5 = 40 STAs are randomly generated. Their channel gains are calculated using ray-tracing methods in MATLAB. We configure the XR video devices with 2 antennas and UL transmit power 18 dBm in the 6 GHz band. All other devices are configured with 2 antennas and UL transmit power 18 dBm in the 5 GHz band.\par

\setlength{\textfloatsep}{5pt}

\begin{figure*}[!t]
  \centering
  \subfloat[XR video]{%
    \includegraphics[width=0.333\textwidth]{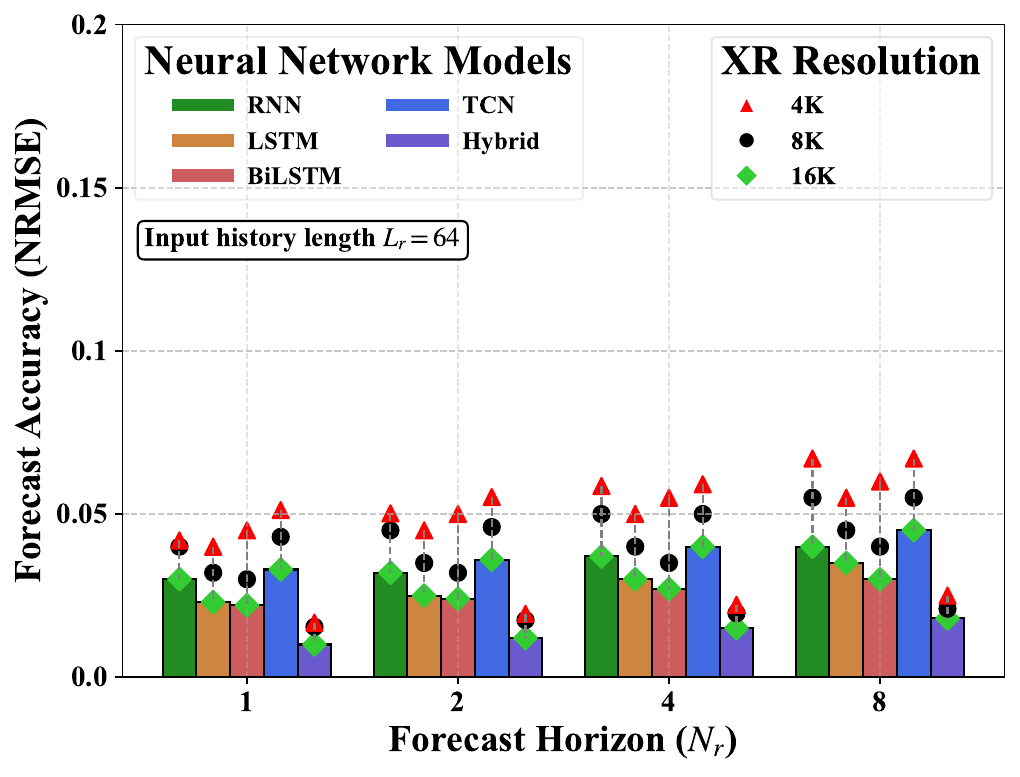}\label{rmsse11}%
  }
  \subfloat[HMD orientations]{%
    \includegraphics[width=0.333\textwidth]{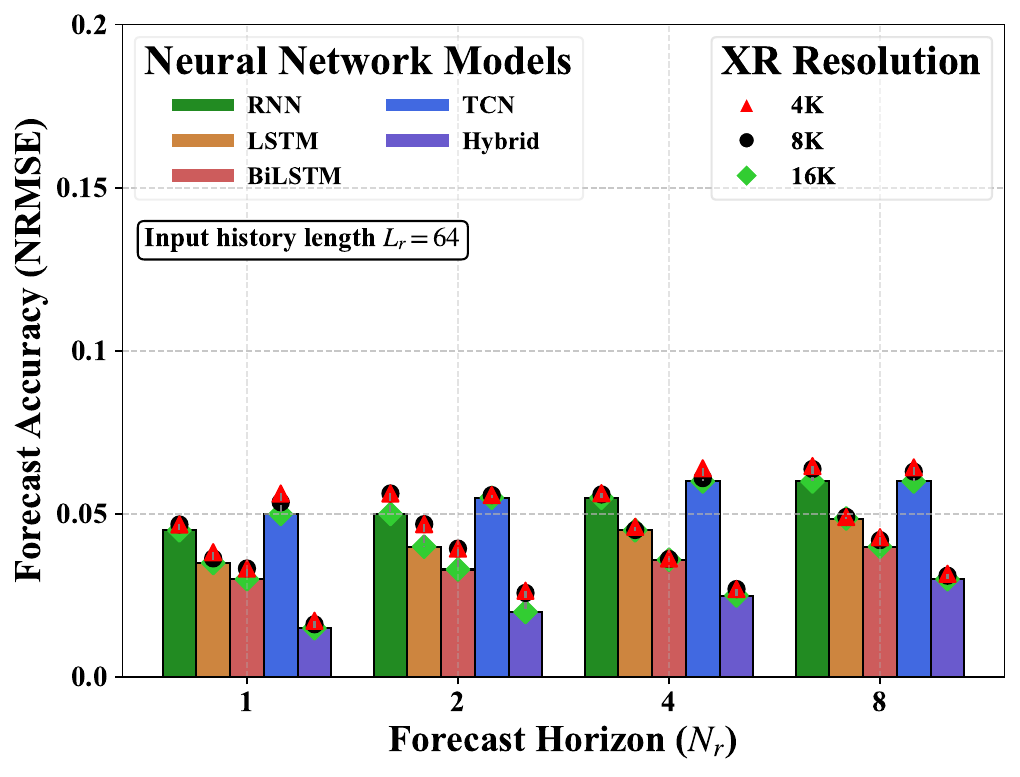}\label{rmsse12}%
  }
  \subfloat[Control and Haptics]{%
    \includegraphics[width=0.333\textwidth]{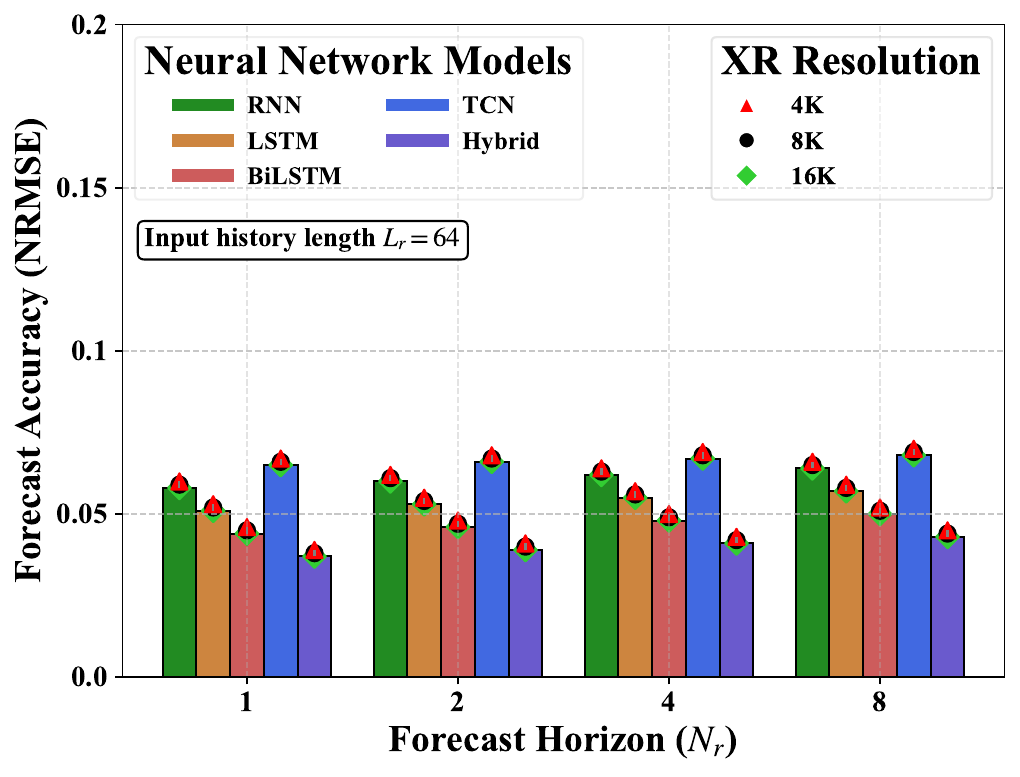}\label{rmsse13}%
  }

  \caption{Comparison of NRMSE of considered time-series forecasts by predictive models with training history length of $L_r = 64$ samples for: (a) XR video, (b) HMD orientations, and (c) control and haptic traffic, against multi-sample-ahead forecast horizons.}
  \label{rmsse_pred1}
\end{figure*}
\setlength{\textfloatsep}{1pt}

\begin{figure*}[!t]
  \centering
  \subfloat[XR video]{%
  \includegraphics[width=0.333\textwidth]{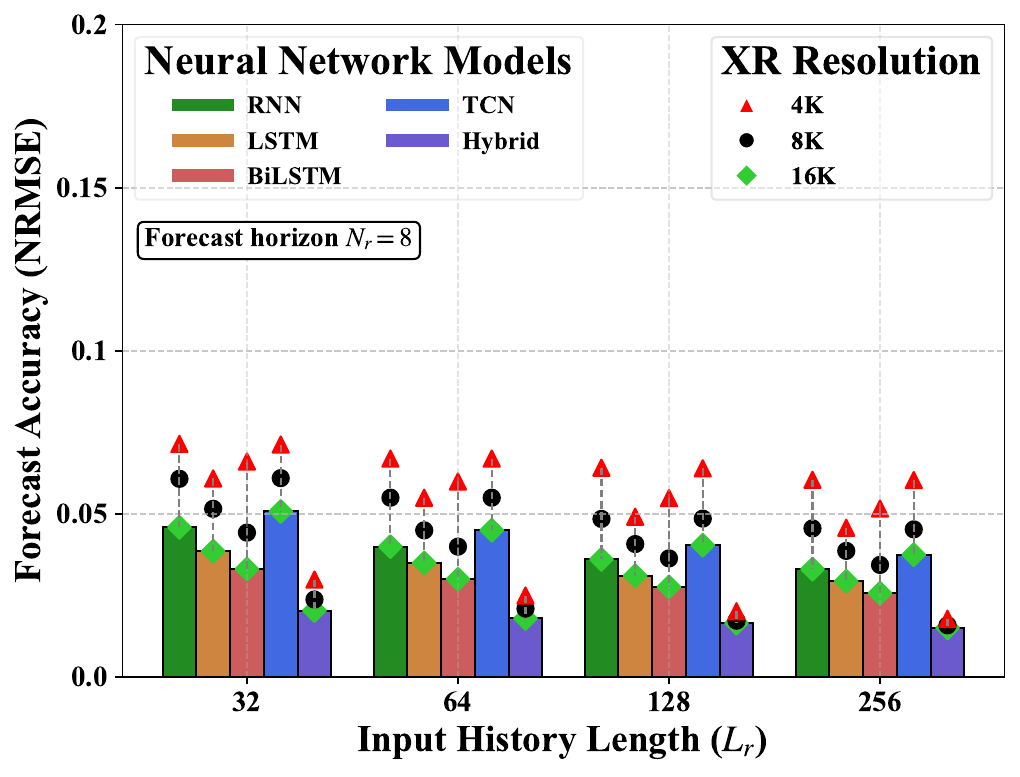}\label{rmsse21}%
  }
  \subfloat[HMD orientations]{%
\includegraphics[width=0.333\textwidth]{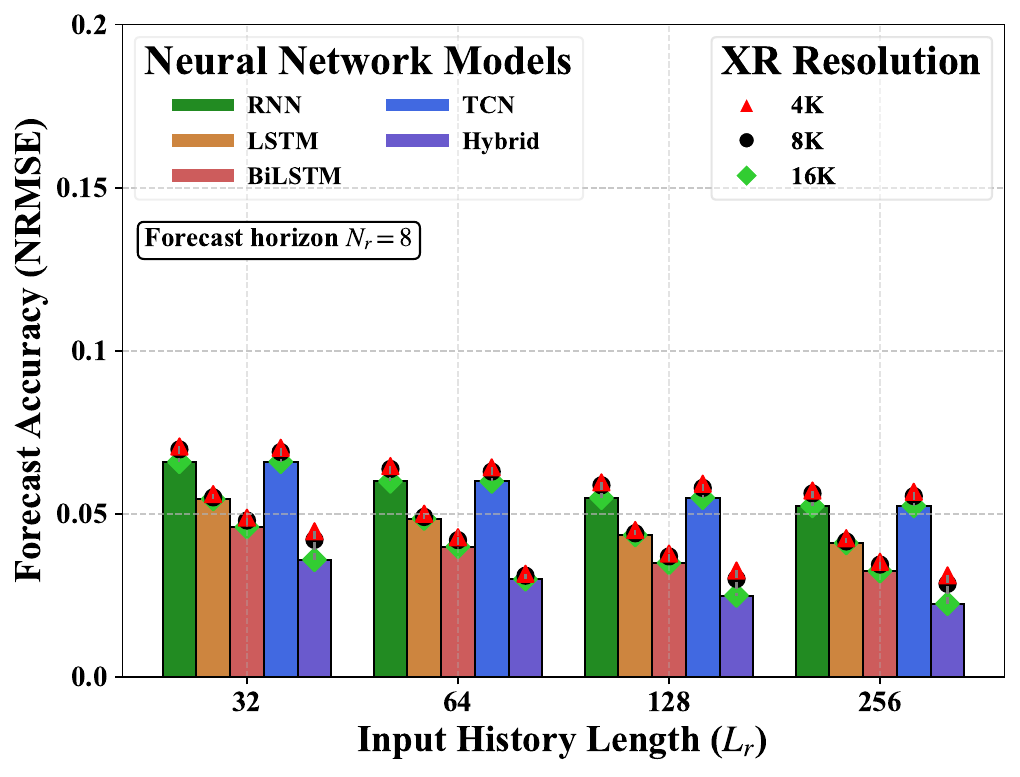}\label{rmsse22}%
  }
  \subfloat[Control and Haptics]{%
\includegraphics[width=0.333\textwidth]{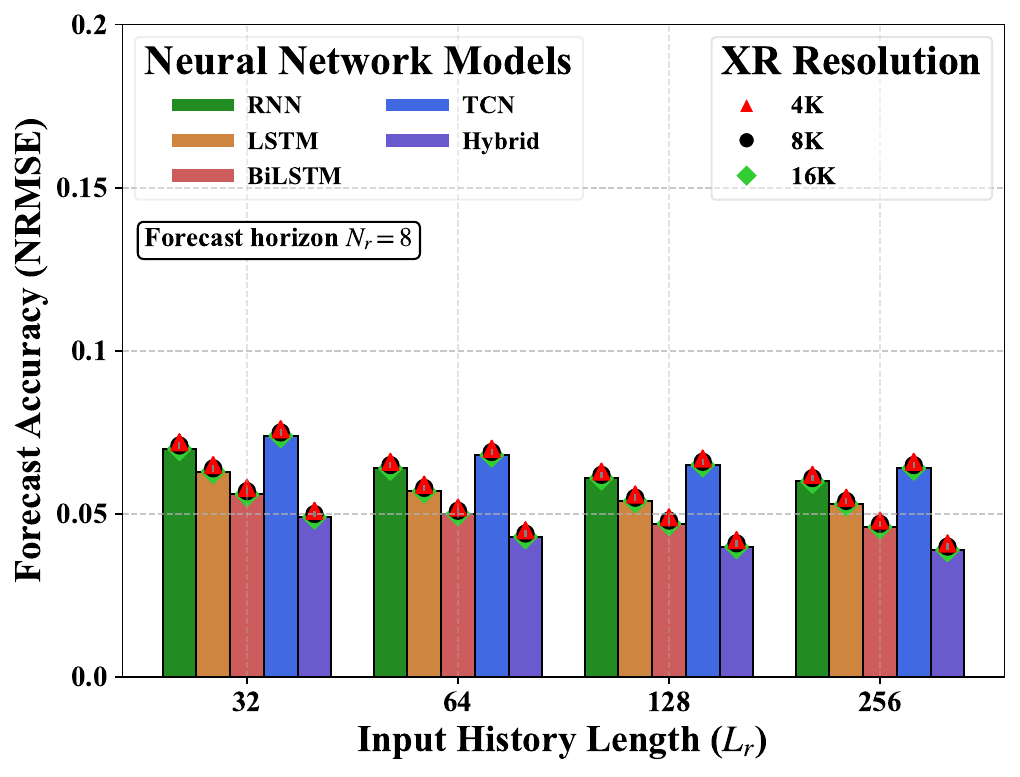}\label{rmsse23}%
  }

  \caption{Comparison of NRMSE of considered time-series forecasts by predictive models with forecast length of $N_r = 8$ samples for: (a) XR video, (b) HMD orientations, and (c) control and haptic traffic, against input history lengths.}
  \label{rmsse_pred2}
\end{figure*}
\setlength{\textfloatsep}{1pt}

\begin{figure*}[!t]
  \centering
  \subfloat[XR video]{\includegraphics[width=0.3\textwidth]{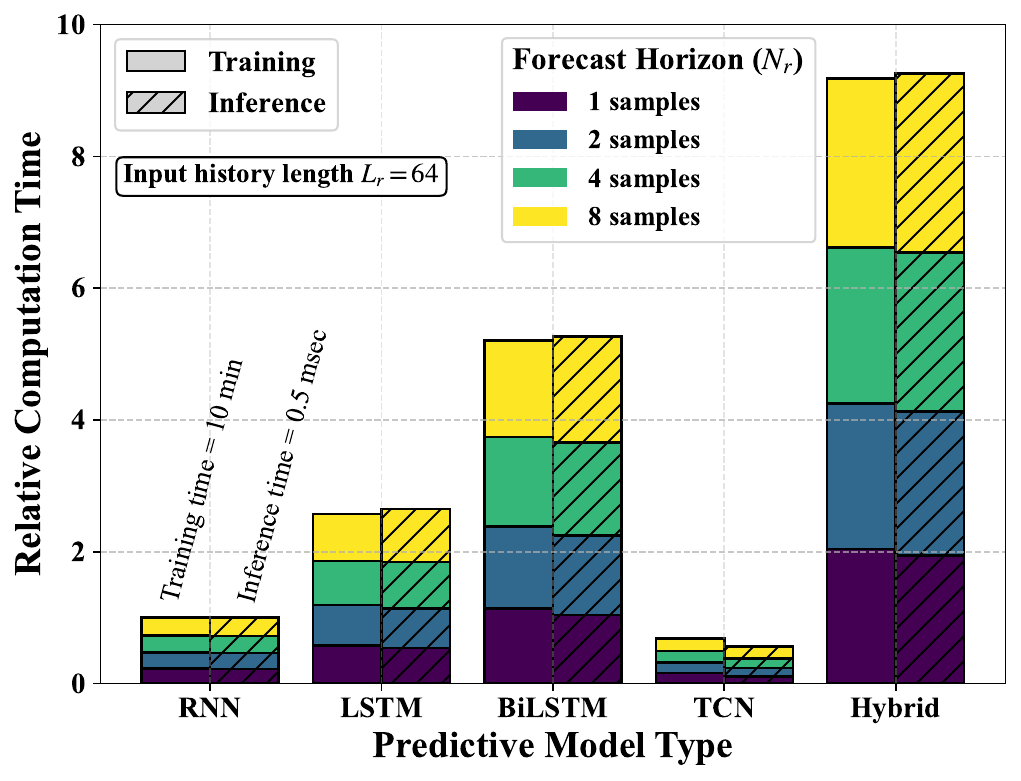}\label{times11}%
  }
  \subfloat[HMD orientations]{%
    \includegraphics[width=0.3\textwidth]{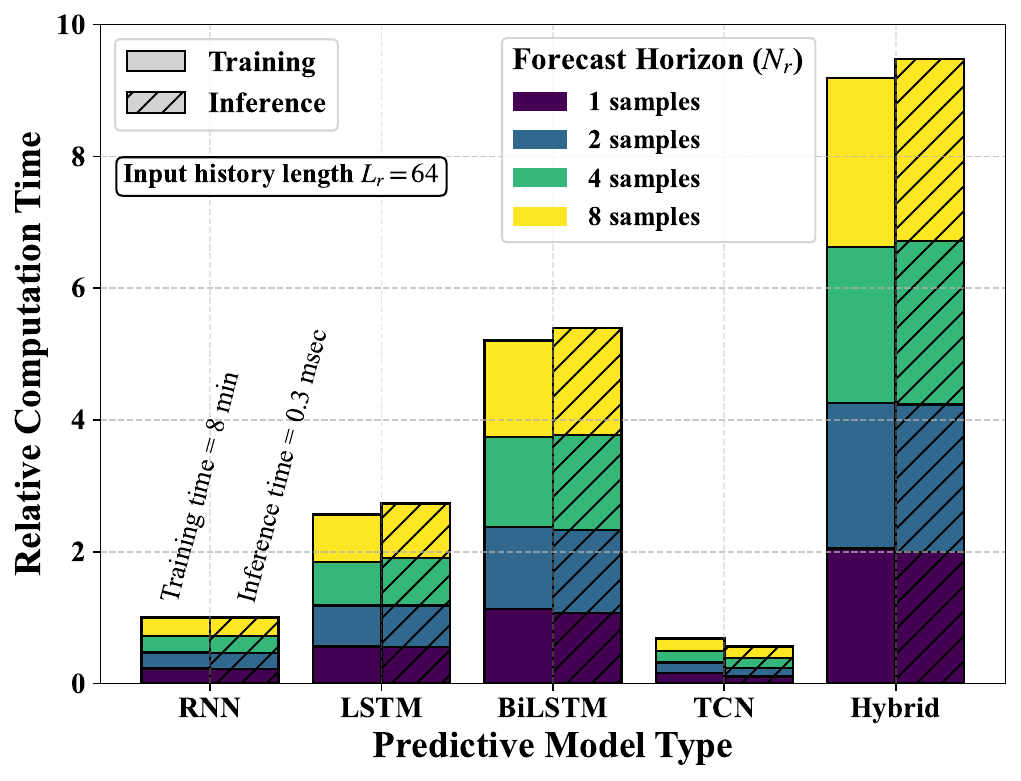}\label{times12}%
  }
  \subfloat[Control and Haptics]{%
    \includegraphics[width=0.3\textwidth]{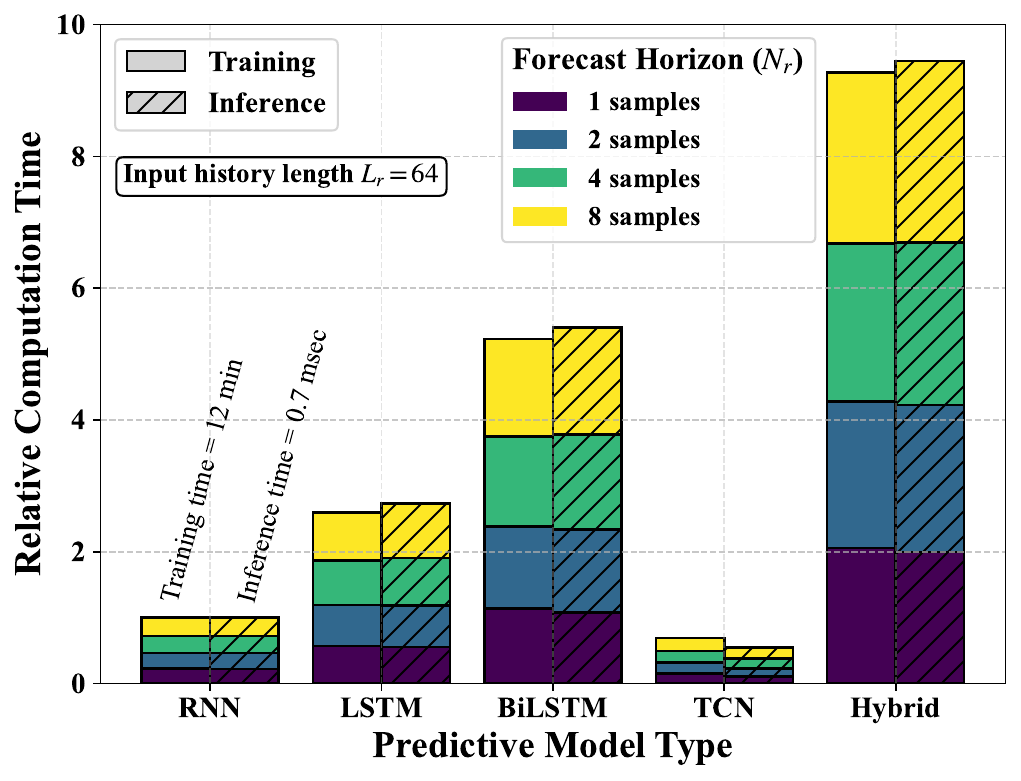}\label{times13}%
  }

  \caption{Comparison of training and inference times of considered time-series predictive models for long-horizon multi-sample-ahead forecasts of: (a) XR video, (b) HMD orientations, and (c) control and haptic traffic.}
  \label{times_pred}
\end{figure*}
\setlength{\textfloatsep}{1pt}


\subsection{Results and Discussions} \label{sec6.2}
The OMNeT++ network simulation was executed for 5 seconds per load instance, with the aggregated network load varying from 0.1 to 1.0. This setup facilitated the collection of traffic arrival data at every polling cycle of 125~$\mu$sec from 256 SFUs ($\sim100\times10^6$ samples) and 32 ONUs ($\sim12\times10^6$ samples) with each of 4K, 8K, and 16K XR video qualities. All generative and predictive traffic forecasting models were implemented using the PyTorch framework, with training and testing conducted on the Spartan high-performance computing cluster to leverage its extensive resources. Following offline training, the models were integrated into an interactive Python$\leftrightarrow$OMNeT++ co-simulation environment to evaluate the performance of the proactive resource allocation schemes.\par
\begin{figure*}[!t]
  \centering
  \subfloat[input history length $L_h$= 128]{%
    \includegraphics[width=0.333\textwidth]{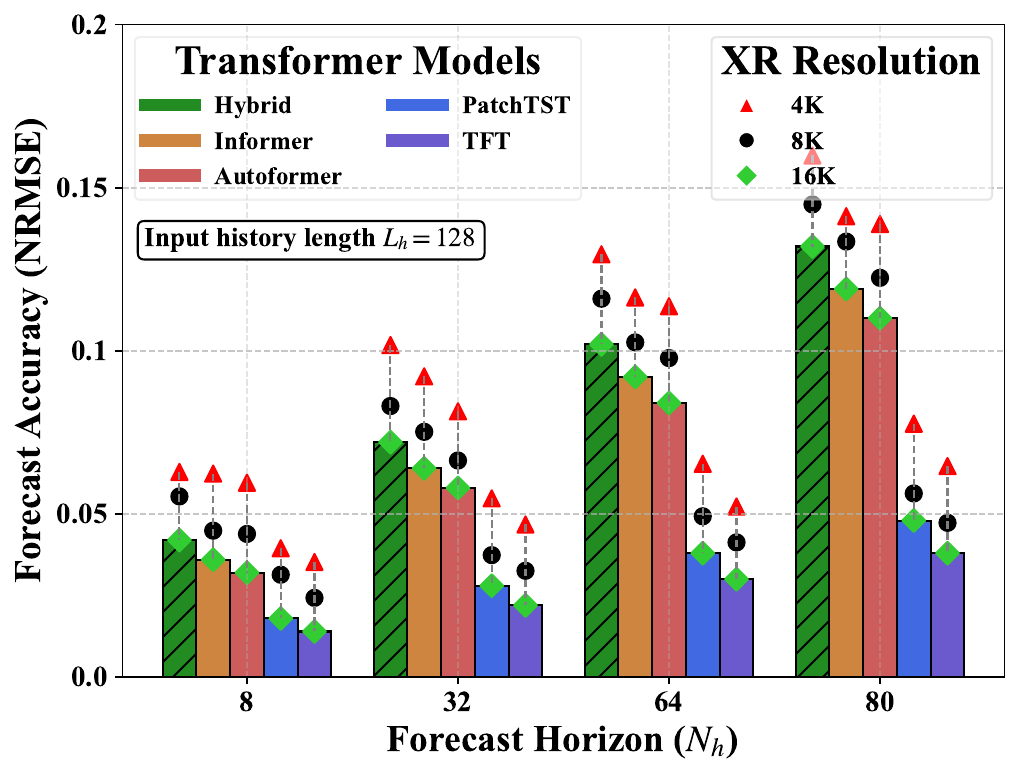}\label{rmsse31}%
  }
  \subfloat[input history length $L_h$ = 256]{%
    \includegraphics[width=0.333\textwidth]{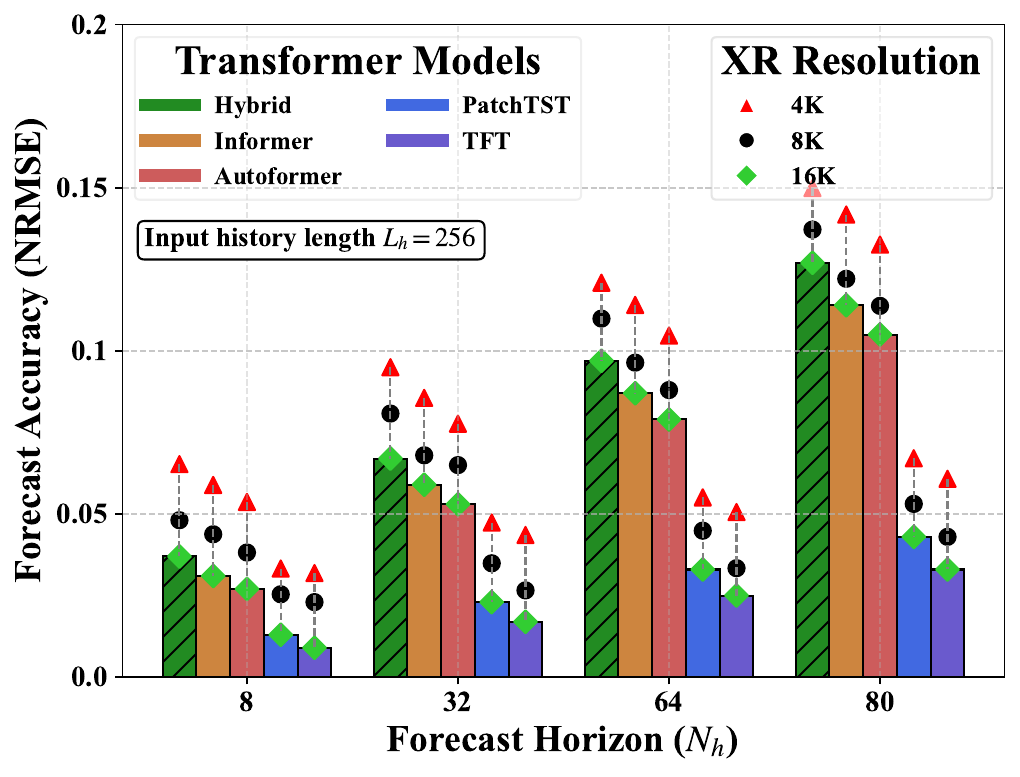}\label{rmsse32}%
  }
  \subfloat[input history length $L_h$ = 512]{%
    \includegraphics[width=0.333\textwidth]{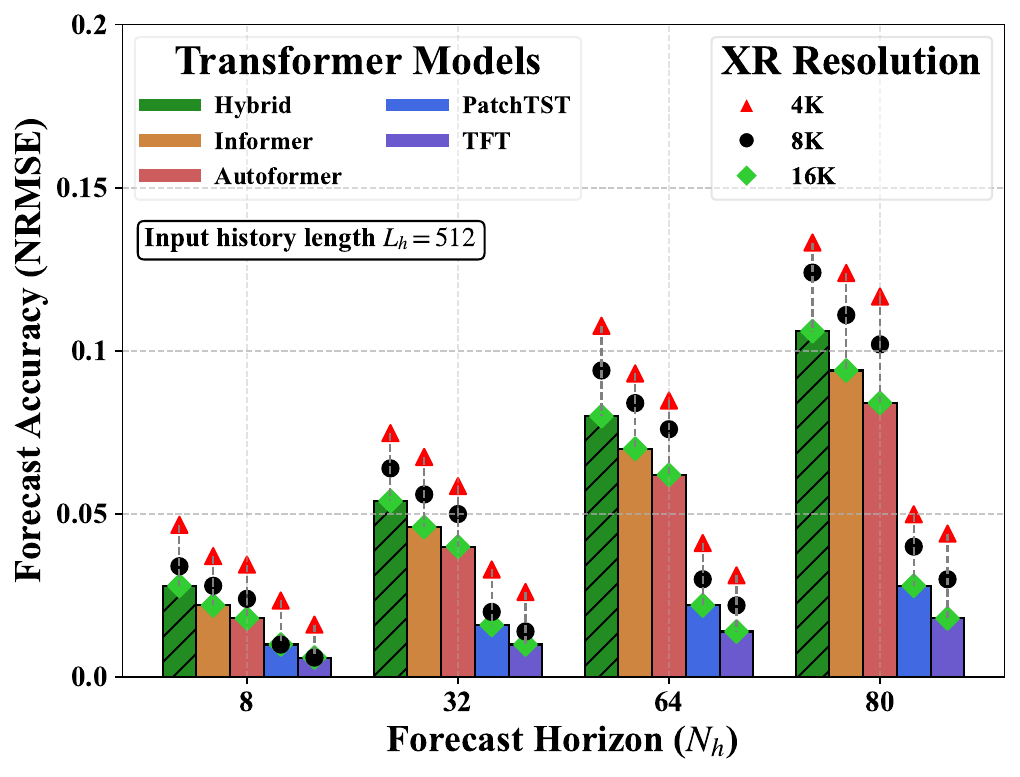}\label{rmsse33}%
  }

  \caption{Comparison of NRMSE of the time-series generative models and the hybrid predictive model against long-horizon forecasts of aggregated XR traffic trained with: (a) history length = 128, (b) history length = 256, and (c) history length = 512.}
  \label{rmsse_gen}
\end{figure*}
\setlength{\textfloatsep}{1pt}

\begin{figure*}[!t]
  \centering
  \subfloat[input history length $L_h$ = 128]{%
    \includegraphics[width=0.333\textwidth]{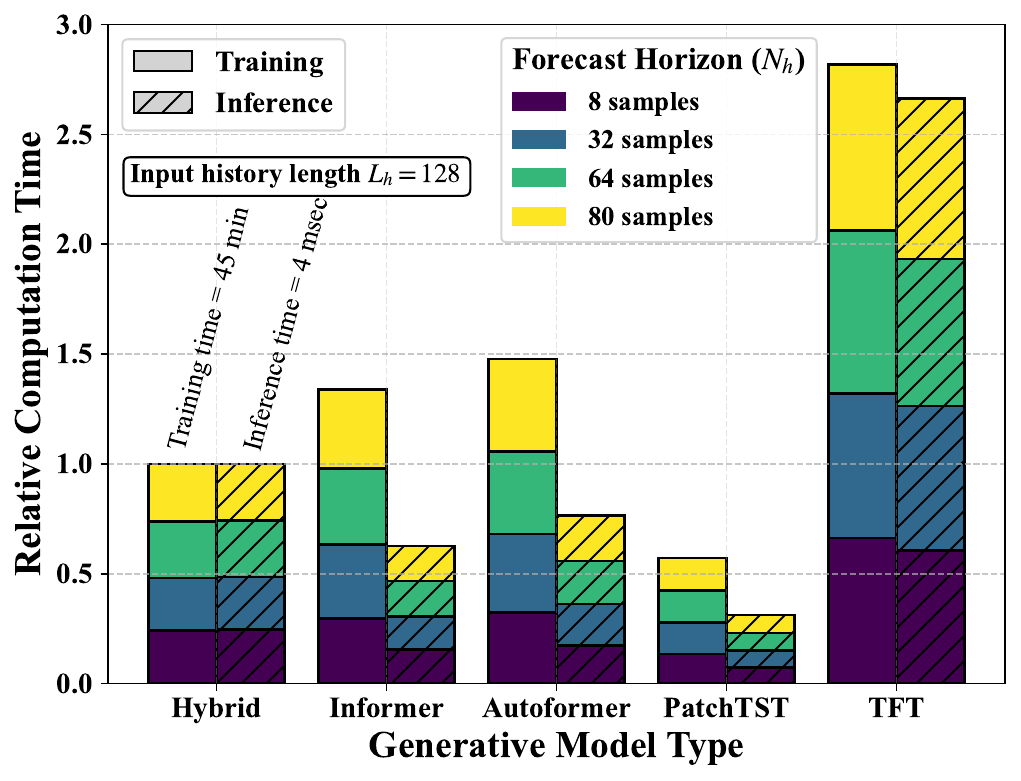}\label{times21}%
  }
  \subfloat[input history length $L_h$ = 256]{%
    \includegraphics[width=0.333\textwidth]{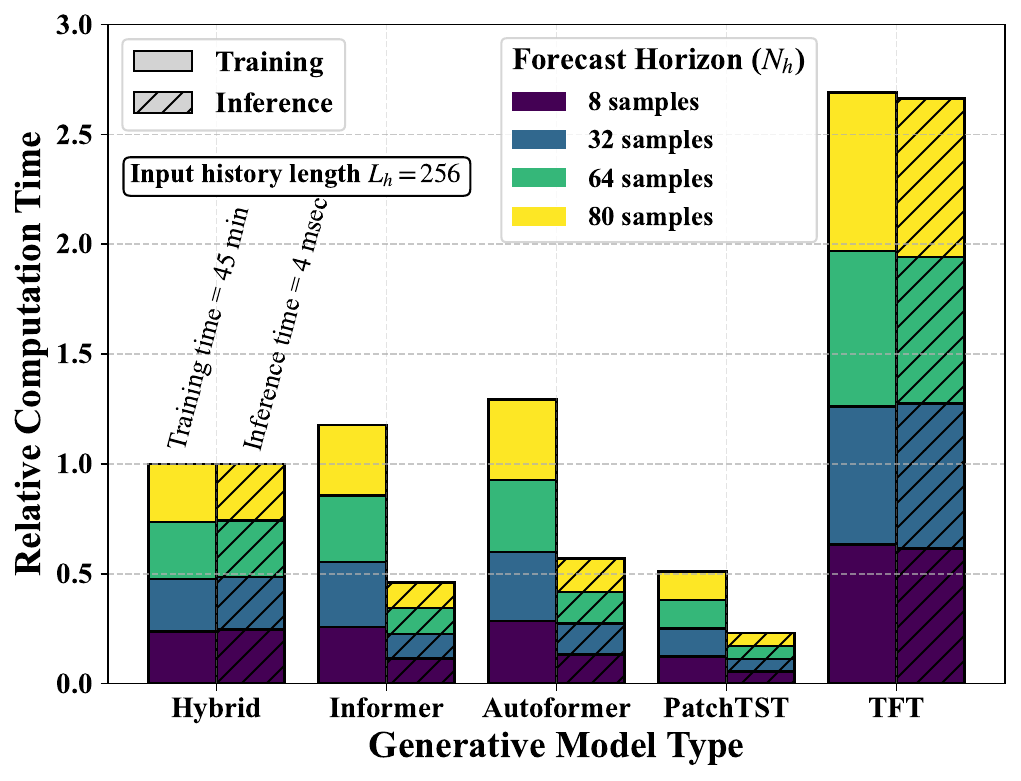}\label{times22}%
  }
  \subfloat[input history length $L_h$ = 512]{%
    \includegraphics[width=0.333\textwidth]{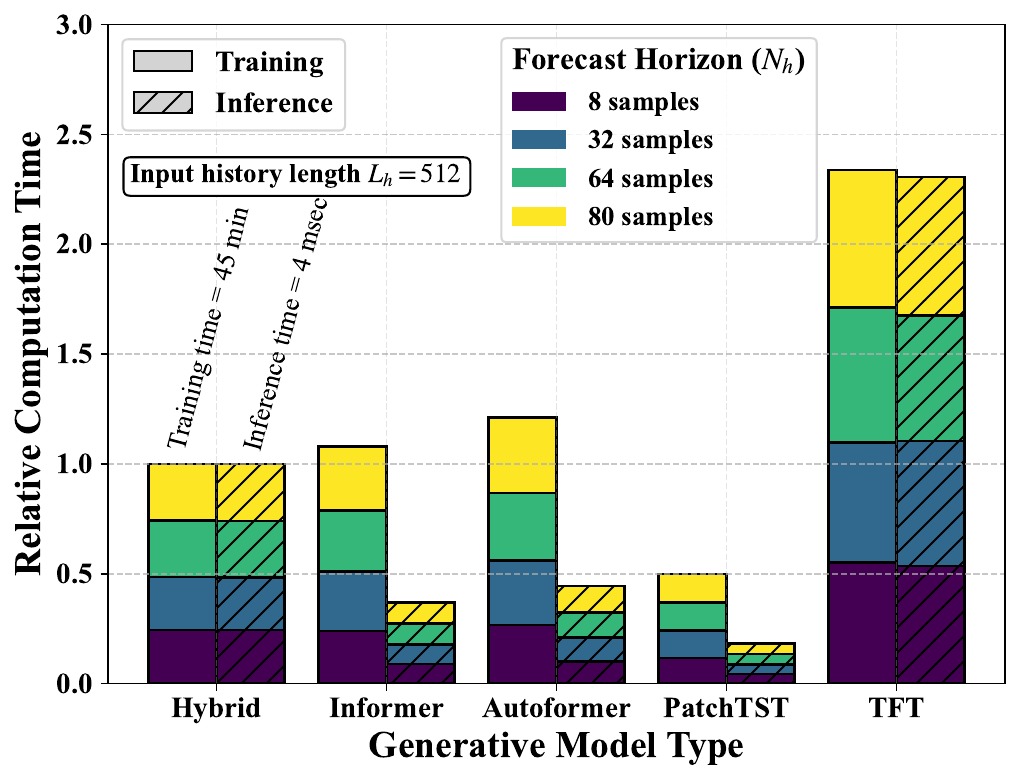}\label{times23}%
  }

  \caption{Comparison of training and inference times of the time-series generative models and the hybrid predictive model for long-horizon forecasts of aggregated XR traffic trained with: (a) history length = 128, (b) history length = 256, and (c) history length = 512.}
  \label{times_gen}
\end{figure*}
\setlength{\textfloatsep}{1pt}
%
Fig.~\ref{rmsse_pred1} shows the NRMSE variation for the considered predictive models, i.e., RNN, LSTM, BiLSTM, TCN, and hybrid (1TCN+2BiLSTM) across long-horizon multi-sample-ahead forecasting of XR video, HMD orientations, and Control/Haptics traffic. With a fixed history length of $L_r$ = 64, we observe a monotonic increase in NRMSE as the forecast horizon extends from $N_r$ = 1 to 8 (125~$\mu$sec to 1 msec), an expected consequence of error accumulation. While the standard RNN is constrained by the vanishing gradient problem, the proposed hybrid architecture achieves significantly lower error rates than all baselines, including the standalone TCN. This accuracy is attributed to the TCN layer effectively extracting local short-term features, which are subsequently processed by the stacked BiLSTM layers to capture temporal dependencies. Note that higher resolution XR video (8K and 16K) result in lower relative errors compared to 4K due to denser, more predictable traffic patterns, whereas the Control and Haptics traffic demonstrate the highest error due to jitter from Generalized Pareto distributed packet arrival times. These observations highlight the importance of architecture design in accommodating both smooth high-rate traffic and stochastic low-rate control streams.\par
Subsequently, we analyze the impact of input history length and computational complexity on these predictive models. As shown in Fig.~\ref{rmsse_pred2}, increasing the history length $L_r$ generally decreases NRMSE by capturing long-term dependencies and improving robustness against bursts, but extending beyond $L_r$ = 64 yields minimal gains. In terms of complexity, the sequential nature of RNN, LSTM, BiLSTM, and hybrid models results in significantly higher training and inference times compared to the TCN, as shown in Fig.~\ref{times_pred}. Although the hybrid model demands more offline training resources, its superior accuracy justifies the cost. To ensure practical deployment, the starting times of model inference must be adjusted according to Section 5.A to effectively embed them within the FTTR-WiFi proactive allocation schemes.\par
%
Next, Fig.~\ref{rmsse_gen} shows the  forecasting accuracy of aggregated multi-modal XR traffic at the ONUs using the considered Transformer-based generative models, i.e., Informer, Autoformer, PatchTST, and TFT alongside the best predictive hybrid model. As shown in Figs.~\ref{rmsse31}-\ref{rmsse33}, both Informer and Autoformer outperform the hybrid model over long horizons $N_h \in \{8,32,64,80\}$, with PatchTST and TFT achieving further notable ($\sim$75\%) improvements. The hybrid model suffers from sequential error accumulation, whereas generative models exploit self-attention to capture global temporal structure and benefit from the smoother aggregated waveforms arising from statistical multiplexing. These results confirm that the generative models provide more stable long-horizon multi-sample-ahead forecasts and substantially lower NRMSE, particularly at longer horizons where the accuracy of sequential models degrade more rapidly.\par
A comparison of the computational cost of the generative models relative to the hybrid model is shown in Fig.~\ref{times_gen}. Due to global self-attention, most generative models exhibit higher training times, with PatchTST being the primary exception as its patching mechanism reduces sequence length and overall complexity. In contrast, inference latencies for Informer, Autoformer, and PatchTST are generally lower than the hybrid model because Transformer layers allow parallel evaluation, making them suitable for real-time predictive scheduling. The only outlier is TFT, whose variable-selection and gating networks introduce substantial overhead that results in inference times exceeding those of the hybrid model. This distinction highlights that while Transformers offer strong accuracy-latency trade-offs, not all variants meet stringent real-time constraints, especially when additional interpretability modules are included like in TFT.\par
\begin{figure*}[!t]
  \centering
  \subfloat[Objective and computation time comparison]{%
    \includegraphics[width=0.333\textwidth]{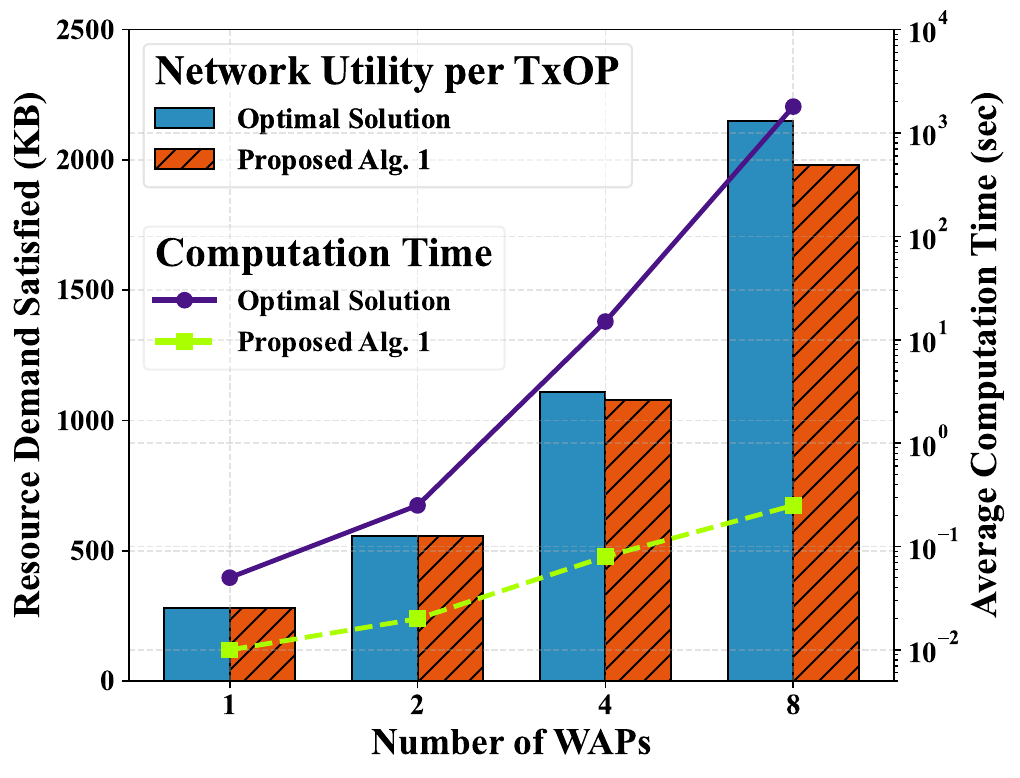}
    \label{wifi1}%
  }
  \subfloat[Wireless interface latency comparison]{%
    \includegraphics[width=0.333\textwidth]{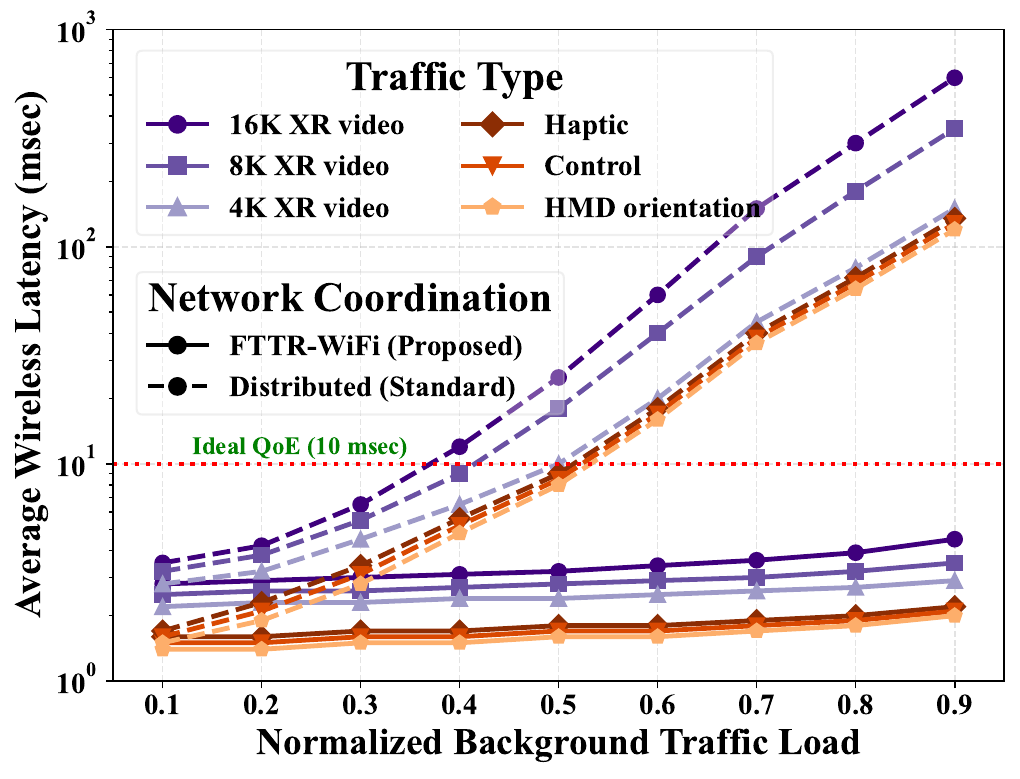}\label{wifi2}%
  }
  \subfloat[Wireless interface latency reduction by handover]{%
    \includegraphics[width=0.333\textwidth]{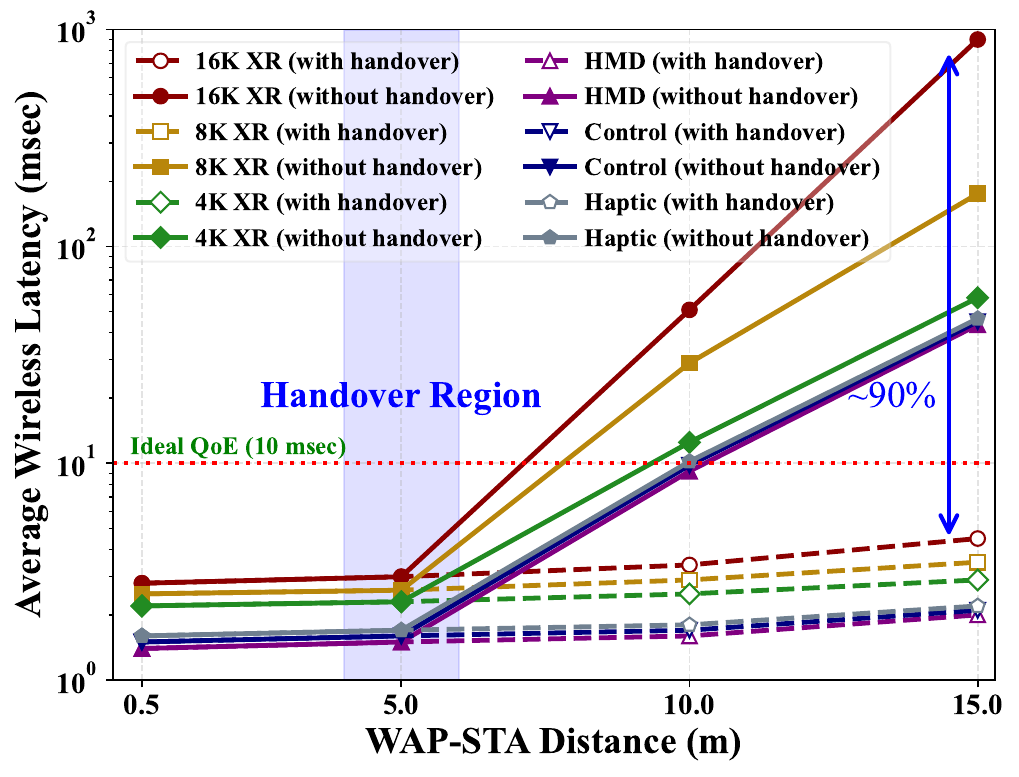}\label{wifi3}%
  }

  \caption{Comparison of: (a) network utilities achieved and computation times consumed by the optimal solution and the proposed Alg.~\ref{alg1}, (b) average wireless interface latency of multi-modal XR traffic with the proposed FTTR-WiFi coordination and standard distributed WiFi, and (c) average wireless interface latency reduction with the proposed network-orchestrated intelligent in-premise handovers.}
  \label{wifi_perf}
\end{figure*}

Efficient STA-WAP associations and seamless in-premise handovers are essential for meeting the end-to-end QoE of multi-user XR.  Fig.~\ref{wifi1} demonstrates that Alg.~\ref{alg1} approximates the optimal objective function value or the network utility obtained via Gurobi solver with substantially lower computational time, which is critical for adapting to dynamic mobility. In Fig.~\ref{wifi2}, the proposed FTTR-WiFi coordinated framework is shown to consistently maintains wireless interface latency within ideal QoE requirements while distributed WiFi networks without coordination is unable to satisfy to these strict QoE requirements. Furthermore, as users move between rooms, the increasing WAP-STA distance degrades channel gain and throughput. Hence, handing over to a neighboring WAP using the proposed strategy significantly mitigates this degradation.  This is evidenced in Fig.~\ref{wifi3} where 16K XR frame latency is reduced by nearly 90\% compared to when maintaining a connection at 15~meter (Fig.~\ref{wifi3}).\par
\setlength{\textfloatsep}{1pt}

\begin{figure*}[!t]
  \centering
  \subfloat[4K XR video for fair QoE]{%
    \includegraphics[width=0.333\textwidth]{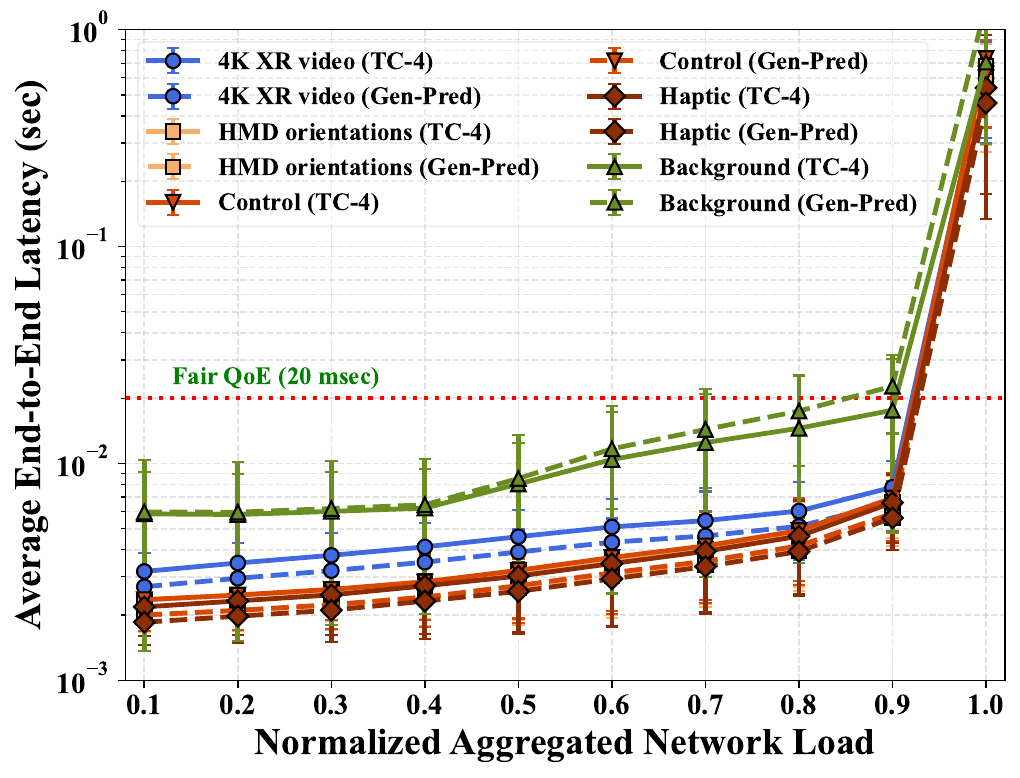}\label{e2e_4k}%
  }
  \subfloat[8K XR video for comfortable QoE]{%
    \includegraphics[width=0.333\textwidth]{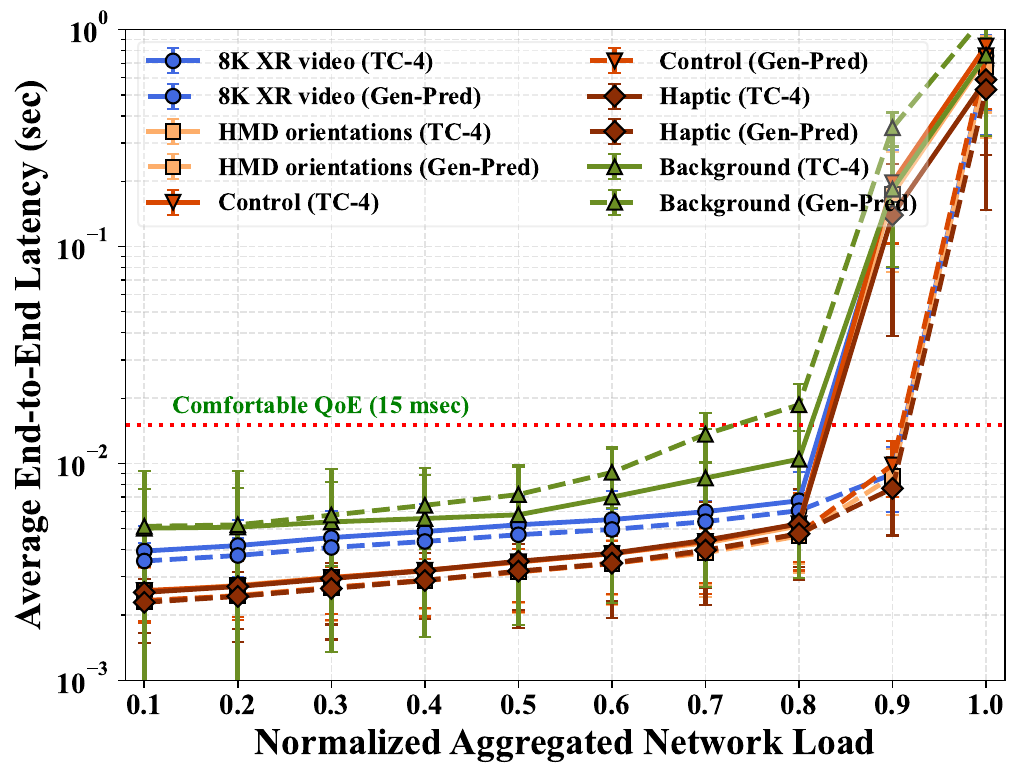}\label{e2e_8k}%
  }
  \subfloat[16K XR video for ideal QoE]{%
    \includegraphics[width=0.333\textwidth]{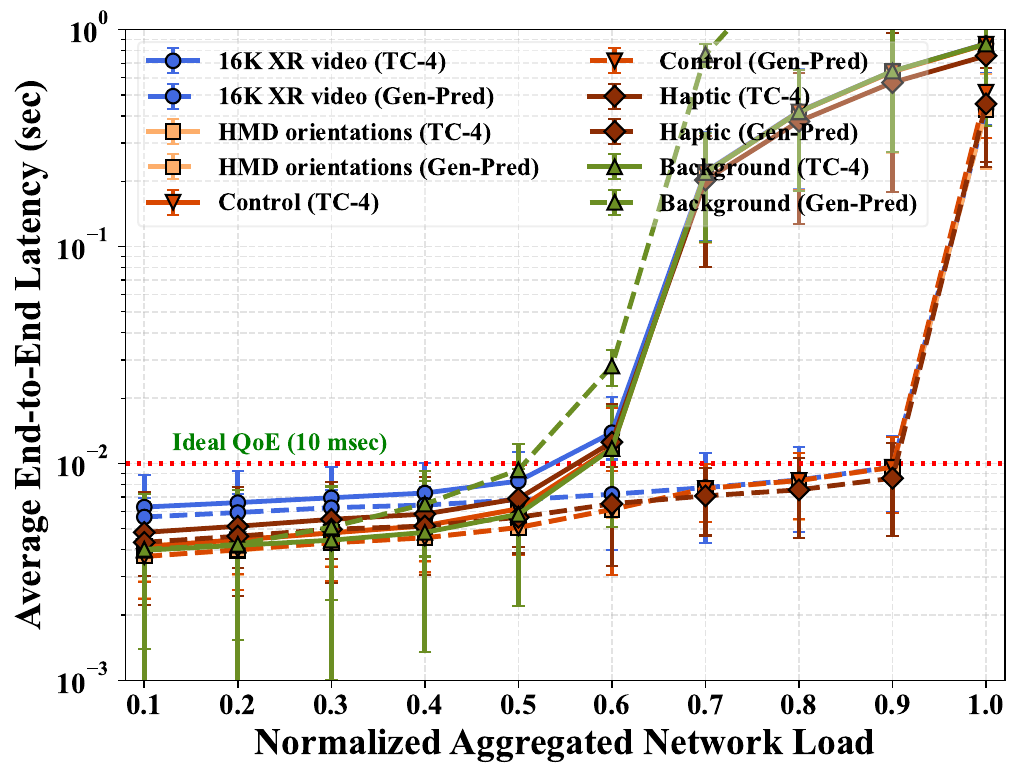}\label{e2e_16k}%
  }
  \caption{Comparison of end-to-end latency of video, HMD orientations, control, haptic, and background traffic in XR collaborations with the proposed Gen-Pred AI resource allocation scheme against G-PON TC-4 baseline scheduling scheme over the integrated FTTH (50G-PON, 1:32), FTTR (XGS-EPON, 1:8), and WiFi-7 while transmitting XR video frames of quality (a) 4K, (b) 8K, and (c) 16K.}
  \label{e2e_latency}
\end{figure*}
\setlength{\textfloatsep}{1pt}
%
Finally, the end-to-end latency of multi-modal XR traffic over the integrated FTTR architecture is presented in Fig.~\ref{e2e_latency}. The average latencies for 4K and 8K XR video under the default reactive TC-4 resource allocation (Figs.~\ref{e2e_4k}-\ref{e2e_8k}) remain within fair and comfortable QoE requirements, respectively. However, they exhibit significant jitter which degrades the user experience. Conversely, as illustrated in Fig.~\ref{e2e_16k}, the latency for 16K video violates the ideal QoE requirement once the normalized aggregated network load surpasses 0.5. Nevertheless, by adopting the proposed Gen-Pred AI resource allocation framework, the end-to-end latency is minimized and jitter is substantially suppressed across all XR qualities. Crucially, even for 16K XR under heavy load conditions, both average delay and jitter are maintained well within the ideal QoE requirements. These results conclusively establish that efficient proactive resource allocation is indispensable for satisfying the high-throughput, ultra-low-latency, and high-reliability requirements of immersive XR collaborations.\par

\section{FROM INFRASTRUCTURE TO SERVICE PLATFORM}
Beyond connectivity, new FTTR rollouts can be monetized by operators providing value-added services. One emerging service model is the FTTR + "x” \cite{SunFTTRX2024,HuaweiFTTRX2025}, in which connectivity is bundled with additional "x" service(s) including:

\textbf{Digital Entertainment Services:} Premium digital entertainment, including UHD video, cloud gaming and immersive XR services, represent an obvious example. As already discussed in Sections 4 and 5, the fiber infrastructure provides capacity while the centrally-managed architecture allows for efficient coordination of resources, thus satisfying latency, consistency and reliability requirements.

\textbf{Integrated Smart-Home:} A second opportunity is an integrated smart-home service architecture. Existing smart-home devices including surveillance cameras, smart locks, environmental sensors and appliances—typically operate through heterogeneous protocols, applications and cloud platforms. In FTTR, the MFU provides a natural aggregation and control point for traffic originating from devices associated with different SFU-WAPs. Service orchestration, device management and traffic policies can therefore be implemented centrally rather than independently at each WAP. In this model, FTTR provides not only connectivity but also a common service plane through which operators can manage heterogeneous devices and data flows across the premises.
    
\textbf{Local Storage Service:} The MFU can additionally incorporate network-attached storage (NAS), transforming the residential gateway into a local storage and content-distribution node. User-generated photographs, video and documents can be stored within the premises and accessed through the high-speed fiber network, while selected content can be synchronized with external cloud storage for resilience and backup. This architecture supports a hybrid edge–home–cloud storage model, in which latency-sensitive or frequently accessed content remains local while less frequently accessed data can reside remotely. By adding AI inference and local computing capability at the MFU, traffic and content-access statistics can further be analyzed to identify frequently requested objects for local caching. For high-bandwidth services such as UHD and immersive XR applications, local caching reduces upstream traffic and content retrieval latency while improving service responsiveness.
    
\textbf{AI-Native Services:} Embedding AI inference and local computing capability in the MFU further expands the return-on-investment of FTTR. As discussed in the preceding paragraph, machine intelligence at the MFU can exploit collected traffic, channel and device-state information to perform various value-add in-premise functions.  Apart from intelligent resource management and content caching, another example is in WiFi sensing.  WiFi sensing can exploit variations in channel-state information and other radio measurements caused by human movement and changes in the physical environment. When sensing information from multiple SFU-WAPs is aggregated at the MFU, local inference could support applications such as occupancy detection, activity recognition, fall detection and anomalous-behavior detection. Importantly, computation can be performed within the premises rather than continuously transferring raw sensing data to a remote cloud. This reduces transport latency and upstream bandwidth requirements, while enabling sensitive sensing data to remain locally processed where appropriate, thereby enhancing data privacy and sovereignty.

\section{Conclusions} \label{sec7}
FTTR establishes a high-capacity, low-latency, and centrally managed optical infrastructure throughout the premises that complements the mobility provided by advanced wireless networks. In this article, we examined the multi-modal traffic of immersive XR applications, the QoE requirements and their implications on network performance.  To address user-experience requirements, we presented a novel framework for proactive coarse-grained resource allocation and intelligent roaming across the integrated FTTR network architecture. 
By incorporating accurate traffic forecasts into the proactive resource allocation framework, results show that high-throughput, low-latency, and high-reliability requirements for XR collaborations can be satisfied, even under high network load conditions which otherwise cannot be achieved with current resource allocation schemes.  Moreover, utilizing the proposed network-orchestrated seamless roaming strategy, the XR frame transmission latency is reduced by nearly 90\% compared to maintaining connections at large WAP-STA distances. This ensures the wireless interface latency remains within the ideal QoE requirements. In summary, FTTR creates a foundation upon which next-generation immersive applications can be deployed. Consequently, the value of FTTR extends beyond its ability to transport data, in that it provides a platform for delivering computation, intelligence, sensing, analytics, and immersive applications directly within the user environment.

\bibliography{sample}

\section*{Author Biographies}

\setlength\intextsep{0pt}

   \noindent
   \textbf{Elaine Wong} received B.E. and Ph.D. degrees from the University of Melbourne, Australia. She is currently the Pro Vice-Chancellor (People \& Equity) and Redmond Barry Distinguished Professor at the  University of Melbourne. Her research interests include energy-efficient optical and wireless networks, optical and wireless technologies for immersive applications, and applied machine learning for low-latency networking. She has coauthored more than 150 journal and conference publications. She is an Optica Fellow and has served on the editorial board of the \emph{Journal of Lightwave Technology} and the \emph{Journal of Optical Communications and Networking}.\\

   \noindent
   \textbf {Sourav Mondal} received PhD from the University of Melbourne in 2020, M.Tech from Indian Institute of Technology Kharagpur in 2014, and B.Tech from Kalyani Govt. Engineering College, affiliated to West Bengal University of Technology in 2012. Currently, he is working as a Research Fellow in the Faculty of Engineering and IT, Department of Electrical and Electronic Engineering of the University of Melbourne. Prior to this, he briefly worked as an EDGE/Marie Skłodowska-Curie Post-doctoral Fellow in Trinity College Dublin, Ireland and as a Senior Lead Systems Engineer (Corporate R\&D) in Qualcomm Bengaluru Design Centre, India.\\

\end{document}